\documentclass[twocolumn]{aastex631}
\usepackage{mathtools}

\newcommand{\mum}{${\rm{\mu}}$m}

\shorttitle{SDSS\,1557}
\shortauthors{Amaro et al.}

\begin{document}

\title{Time-resolved Hubble Space Telescope Wide Field Camera 3 Spectrophotometry Reveals Inefficient Day-to-Night Heat Redistribution in the Highly Irradiated Brown Dwarf SDSS\,1557B}

\correspondingauthor{Rachael Amaro}
\email{rcamaro@arizona.edu}

\author[0000-0002-1546-9763]{Rachael C. Amaro}
\altaffiliation{National Science Foundation Graduate Research Fellow}
\affiliation{Department of Astronomy and Steward Observatory, The University of Arizona, 933 North Cherry Avenue, Tucson, AZ 85721, USA}

\author[0000-0003-3714-5855]{D\'aniel Apai}
\affiliation{Lunar and Planetary Laboratory, The University of Arizona, 1629 E. University Blvd, Tucson, AZ 85721, USA}
\affiliation{Department of Astronomy and Steward Observatory, The University of Arizona, 933 North Cherry Avenue, Tucson, AZ 85721, USA}

\author[0000-0003-1487-6452]{Ben W. P. Lew}
\affiliation{Bay Area Environmental Research Institute and NASA Ames Research Center, Moffett Field, CA 94035, USA}

\author[0000-0003-2969-6040]{Yifan Zhou}
\affiliation{Department of Astronomy, 530 McCormick Rd, Charlottesville, VA 22904, USA}

\author[0000-0003-3667-8633]{Joshua D. Lothringer}
\affiliation{3700 San Martin Drive, Baltimore, MD 21218}

\author[0000-0003-2478-0120]{Sarah L. Casewell}
\altaffiliation{STFC Ernest Rutherford Fellow}
\affiliation{School of Physics and Astronomy, University of Leicester, Leicester LE1 7RH, UK}

\author[0000-0003-2278-6932]{Xianyu Tan}
\affiliation{Tsung-Dao Lee Institute, Shanghai Jiao Tong University, 520 Shengrong Road, Shanghai, People’s Republic of China}
\affiliation{School of Physics and Astronomy, Shanghai Jiao Tong University, 800 Dongchuan Road, Shanghai, People’s Republic of China}

\author[0000-0002-7129-3002]{Travis Barman}
\affiliation{Lunar and Planetary Laboratory, The University of Arizona, 1629 E. University Blvd, Tucson, AZ 85721, USA}

\author[0000-0002-5251-2943]{Mark S. Marley}
\affiliation{Lunar and Planetary Laboratory, The University of Arizona, 1629 E. University Blvd, Tucson, AZ 85721, USA}

\author[0000-0002-4321-4581]{L. C. Mayorga}
\affiliation{The Johns Hopkins University Applied Physics Laboratory, 11100 Johns Hopkins Rd, Laurel, MD, 20723, USA}

\author[0000-0001-9521-6258]{Vivien Parmentier}
\affiliation{Laboratoire Lagrange, Observatoire de la C\^ote d’Azur, Universit\'e C\^ote d’Azur, Nice, France}

\begin{abstract}
Brown dwarfs in ultra-short period orbits around white dwarfs offer a unique opportunity to study the properties of tidally-locked, fast rotating (1--3~hr), and highly-irradiated atmospheres. Here, we present phase-resolved spectrophotometry of the white dwarf-brown dwarf (WD--BD) binary SDSS\,1557, which is the fifth WD--BD binary in our six-object sample. Using the Hubble Space Telescope Wide Field Camera 3 Near-infrared G141 instrument, the 1.1 to 1.7 \mum{} phase curves show rotational modulations with semi-amplitudes of 10.5$\pm$0.1\%. We observe a wavelength dependent amplitude, with longer wavelengths producing larger amplitudes, while no wavelength dependent phase shifts were identified. The phase-resolved extracted BD spectra exhibit steep slopes and are nearly featureless. A simple radiative energy redistribution atmospheric model recreates the hemisphere integrated brightness temperatures at three distinct phases and finds evidence for weak redistribution efficiency. Our model also predicts a higher inclination than previously published. We find that SDSS\,1557B, the second most irradiated BD in our sample, is likely dominated by clouds on the night side, whereas the featureless day side spectrum is likely dominated by H$^-$ opacity and a temperature inversion, much like the other highly-irradiated BD EPIC2122B.

\end{abstract}


%%%%%%%%%%%%%%%%%%%%%%%%%%%%%%%%%%%%%%%%%%%%%%%%%%%%%%%%%%%%%%%%%%%%
%%%%%%%%%%%%%%%%%%%%%%%%%%%%%%%%%%%%%%%%%%%%%%%%%%%%%%%%%%%%%%%%%%%%
\section{Introduction} \label{sec:intro}

Irradiated planetary atmospheres -- most prominently hot Jupiters -- are generally shaped by four major processes: (1) Energy transport via radiation, convection, and large-scale circulation; (2) Atmospheric equilibrium and non-equilibrium chemistry; (3) Photochemistry, often including UV--driven reactions and dissociation; (4) Micro- and macro-physics of clouds and haze layers. With the steadily growing and increasingly high-quality datasets on hot Jupiters, significant progress has been made in understanding the combined impact of these processes. In particular, forced atmospheric circulation in the tidally locked atmospheres of hot Jupiters has been successfully explored \citep{ShowmanGuillot02, Heng15, Showman20, Lian22}, as well as the formation of some cloud species and their spatial distribution \citep{Wakeford15, Parmentier16, Zhang18, Lee18, Powell19}.

\begin{figure*}
\begin{center}
\includegraphics[width=0.99\textwidth]{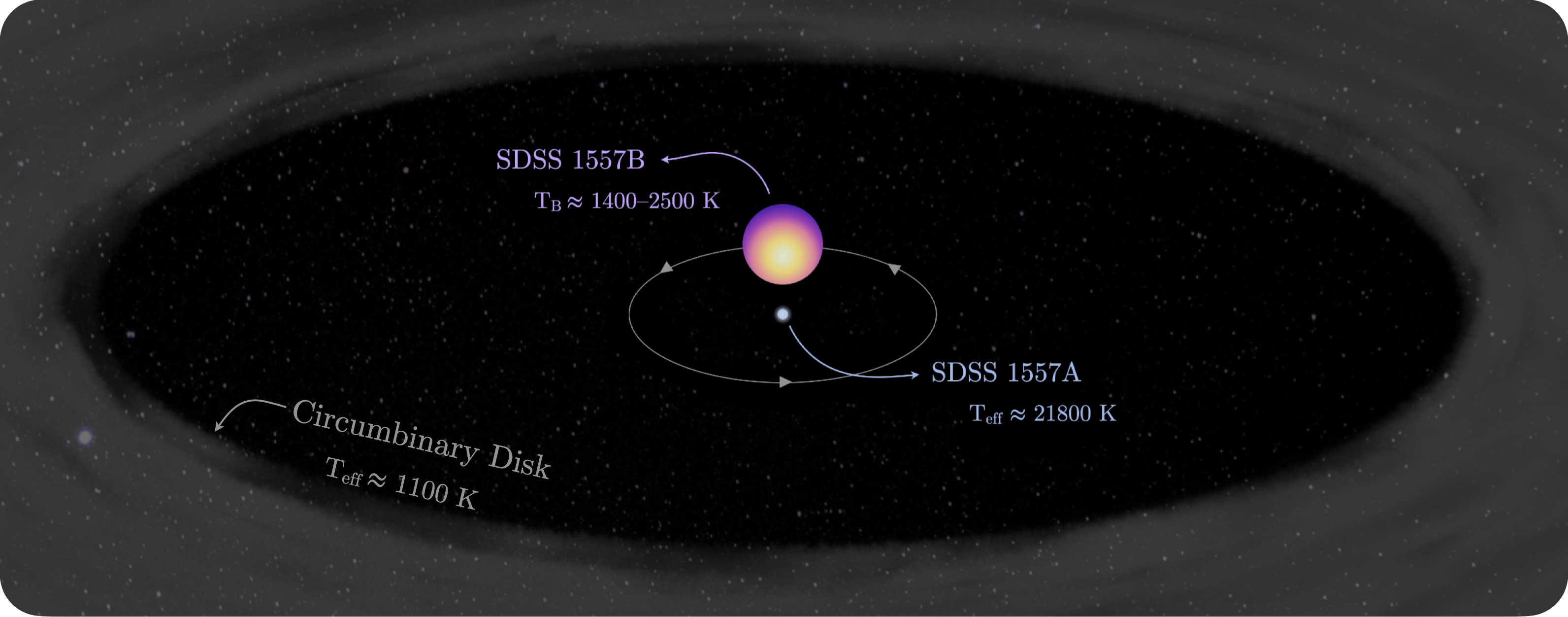}
\caption{Graphic illustration of the SDSS\,1557 system. Radii of the primary white dwarf and secondary brown dwarf are to scale, while orbital distances are not to scale. Colors correspond to calculated temperatures from spectral fitting. The circumbinary dust disk is believed to be accreting onto the WD \citep{Farihi17}, although the accretion is not shown here.}
\label{fig:schematic}
\end{center}
\end{figure*}

However, with any two hot Jupiters differing in multiple key parameters, it remains challenging to unequivocally separate the importance of the above processes and their impact on individual atmospheres. For example, changes in the 3D pressure-temperature distribution and atmospheric chemistry may lead to observational changes that could also be explained by changes in the cloud properties and particle size distributions \citep{Venot20}.

Studies of irradiated atmospheres that significantly differ in rotation and internal heat from hot Jupiters offer important opportunities to better separate the impact of these processes. As pointed out by \citet[][]{Showman20}, highly irradiated brown dwarfs on very short-period (1--3~hr) orbits around white dwarfs (WD--BD systems) offer such targets. 

While only relatively few, close-in WD--BD systems are known today \citep[][]{Casewell12, Casewell20, Beuermann13, Parsons17, Farihi17, Owens23}, due to the range of orbital radii and WD temperatures present in the systems, they offer a broad range of irradiation levels. The irradiance in known systems range from those that correspond to cool hot Jupiters (800~K) \citep{Samland17} to those that correspond to ultra hot Jupiters ($>$2500~K) \citep{Smith11, Sing13}. At the same time, due to the high to very high (7700--25000~K) temperatures of the WDs, a greater fraction of the irradiation is emitted in the ultra-violet than it is the case for hot Jupiters that orbit main sequence stars. Furthermore, the mass of the BDs in WD--BD systems is typically 30--50$\times$ higher than is typical for hot Jupiters, translating to higher gravity atmospheres. Finally, while the orbital and rotational periods are synchronized (just as for hot Jupiters), the orbital periods are about an order of magnitude shorter (1--3~hr). In short, BDs in rare WD--BD systems provide an exciting opportunity to study the same processes that shape hot Jupiters, but to do so in part of the parameter space that overlaps with but also substantially different from those occupied by hot Jupiters.

\begin{deluxetable*}{llcl}
\tablecaption{Properties of the SDSS\,1557 binary system \label{tab:keyprops}}
%\tablewidth{10pt}
\tablehead{
\colhead{Parameter} & \colhead{Units} & \colhead{Value} & \colhead{Reference}
}
%%%%%%%%%%%%%%%%%%%%%%%%%%%%%%%
\startdata
$a$ & Orbital Separation [R$_{\odot}$] & 0.70$\pm$0.02 & \citet{Farihi17} \\
$D$ & Distance [pc] & 500$^{+19.8}_{-18.0}$ & \citet{BailerJones21} \\
$P$ & Orbital Period [mins] & 136.38918$\pm$0.00012 & \citet{Farihi17} \\
$T_{\rm{eff,WD}}$ & Effective Temperature [K] & 21800$\pm$800 & \citet{Farihi17} \\
log$_{10}$($g$) (WD) & Surface Gravity [cm s$^{-2}$] & 7.63$\pm$0.11 & \citet{Farihi17} \\
$M_{\rm{WD}}$ & Mass [M$_{\odot}$] & 0.447$\pm$0.043 & \citet{Farihi17} \\
$M_{\rm{BD}}$ & Mass [M$_{\odot}$] & 0.063$\pm$0.002 & \citet{Farihi17} \\
$R_{\rm{L}}$ & Roche Lobe [R$_{\odot}$] & 0.16 & \citet{Farihi17} \\
Cooling Age (WD) & [Myr] & 33$\pm$5 & \citet{Farihi17} \\
$R_{\rm{WD}}$ & Radius [R$_{\odot}$] & 0.0162 $\pm$ 0.0012 & This work \\
$R_{\rm{BD}}$ & Radius [R$_{\odot}$] & 0.106 $\pm$ 0.024 & This work \\
              & Radius [R$_{\rm{Jup}}$] & 1.054 $\pm$ 0.242 & This work \\
$i$ & Inclination [degrees] & 80 $\pm$ 3 & This work
\enddata
%%%%%%%%%%%%%%%%%%%%%%%%%%%%%%%
\end{deluxetable*}
\vspace{-0.9 cm}
As part of a larger Hubble Space Telescope (HST) program, we observed phase curves of six WD--BD systems using time-resolved G141 infrared spectrophotometry. The high photometric precision ($\sim$0.3\%) achievable with these observations allows us to probe the emission spectra of the brown dwarfs as a function of the longitude (or, equivalently, the rotational phase). Thus, the temporal resolution for these rotating targets provides spatially resolved information on how the atmosphere changes with varying irradiation from the WD host.  Our target is \object{SDSS-J155720.78+091624.7} (hereafter SDSS\,1557), the fifth WD--BD system in our program, and the second-most irradiated. 

This paper is organized as follows: First, in Section~\ref{sec:bdwd_system}, we review the current knowledge on the WD--BD binary target of our study. Next, in Section~\ref{sec:obs_dataredux} we present the new HST time-resolved observations and the data reduction. This is followed by the analysis of the broadband photometric and spectrally dispersed light curves, including model fits (Section~\ref{sec:lcanalysis}). In Section~\ref{sec:bdspec}, we analyze the rotational phase-resolved spectra, including the subtraction of the WD spectrum, extracting spectral modulations as a function of rotational phase, fitting a simple energy balance equation-based temperature distribution model, and also constraining the brightness temperatures in the day and night sides of the BD. In Section~\ref{sec:Compare_models}, we compare the extracted BD spectra to state-of-the-art irradiated atmosphere models. Finally, in Section~\ref{sec:conclusions}, we review the main conclusions of this study.

%%%%%%%%%%%%%%%%%%%%%%%%%%%%%%%%%%%%%%%%%%%%%%%%%%%%%%%%%%%%%%%%%%%%
%%%%%%%%%%%%%%%%%%%%%%%%%%%%%%%%%%%%%%%%%%%%%%%%%%%%%%%%%%%%%%%%%%%%
\section{White Dwarf--Brown Dwarf\\Binary System SDSS\,1557} \label{sec:bdwd_system}

SDSS\,1557 was first suggested to be a WD + debris disk system by \citet{Steele11}, who found Near-Infrared (NIR) flux excesses in the \textit{Y, J}, and \textit{K} photometry from the UKIRT Infrared Deep Sky Survey (UKIDSS) Large Area Survey Data Release 8. Excess \textit{K}--band photometry was best-fit by 700~K blackbody, indicating a potentially cool disc. However, excesses in \textit{Y}-- and \textit{J}--band were best matched by companion of spectral type L4, suggesting a BD companion.

Follow-up observations by \citet{Farihi17} with Gemini+GMOS (2012) and VLT+XSHOOTER (2014-2016) revealed that both sources of NIR excess were likely present within the system, making this the first known WD--BD binary with a metallic circumbinary debris disk. \citet{Farihi17} used radial velocity changes in the Mg\,II 4482~\AA{} absorption feature to reveal a period of 138.4 min, along with a predicted mass ratio of $M_2$/$M_1 = 0.14$ via the H$\alpha$ emission feature and assuming it originated from the companion. Further assuming that the H$\alpha$ emission comes uniformly from the day side hemisphere yields a substellar companion mass of $M_2$ = 0.063 M$_{\rm{\odot}}$ (66 M$_{\rm{Jup}}$) at an orbital inclination of $i\sim$63$^{\circ}$. Note, higher orbital inclination corresponds to more equator-on viewing angles with spin-orbit aligned systems. 

White dwarf effective temperature, $T_{\rm{eff}}$, and surface gravity, log($g$), were derived by \citet{Farihi17} from fitting atmospheric models to Balmer lines from the XSHOOTER spectra. These values, along with photometry, were cross referenced with evolutionary models from \citet{Fontaine01} to predict a cooling age of 33$\pm$5 Myr. Additionally, \citet{Farihi17} found evidence in both the GMOS and XSHOOTER spectra that the metal polluted debris disk was actively accreting onto the WD primary, since detected metal lines such as Mg\,II and Ca\,II K should sink below the photosphere of the WD within a few weeks.

\citet{Swan20} modeled the emission of the SDSS\,1557 system using a combination of two black bodies and found the model overpredicted the observed modulation amplitude at 4.5 \mum{}. They argue that this could be due to flux dilution caused by the dust emission in the debris disk, or caused by molecular absorption and clouds in the companion's atmosphere.

Figure~\ref{fig:schematic} shows a simple illustration of the SDSS\,1557 binary system, along with the debris disk, at an imagined orbital inclination of 63 degrees. The three major bodies are labeled, along with their effective temperatures, or brightness temperatures in the case of the BD. Not illustrated is the process of accretion from the debris disk onto the WD. The WD's color came from \citep{HarreHeller21} and its published effective temperature of 21800~K. The color of the BD came from \citep{Cranmer21} and the estimated day  and night side temperatures derived in this work. Relevant properties this binary system, sourced from literature and this work, can be found in Table~\ref{tab:keyprops}.

\begin{deluxetable}{cllc}
\tablecaption{Log of HST WFC3 Observations for SDSS\,1557\label{tab:obslog}}
\tablewidth{0pt}
\tablehead{
\colhead{Orbit} & \colhead{Observation} & \colhead{Filter} & \colhead{Exp. Start} \\
\colhead{} & \colhead{Type} & \colhead{ID} & \colhead{[BJD$_{\rm{TDB}}$]}
}
%%%%%%%%%%%%%%%%%%%%%%%%%%%%%%%
\startdata
\hline
\multicolumn{4}{c}{Visit 1}\\
\hline
1 & Imaging & F127M & 2459004.53\\
  & Spectroscopic & G141 & 2459004.53\\
2 & Imaging & F127M & 2459004.60\\
  & Spectroscopic & G141 & 2459004.60\\
3 & Imaging & F127M & 2459004.67\\
  & Spectroscopic & G141 & 2459004.67\\
4 & Imaging & F127M & 2459004.73\\
  & Spectroscopic & G141 & 2459004.74\\
\hline
\multicolumn{4}{c}{Inter-visit gap: 995 min}\\
\hline
\multicolumn{4}{c}{Visit 2}\\
\hline
5 & Imaging & F127M & 2459005.46\\
  & Spectroscopic & G141 & 2459005.46\\
6 & Imaging & F127M & 2459005.66\\
  & Spectroscopic & G141 & 2459005.66\\
7 & Imaging & F127M & 2459005.72\\
  & Spectroscopic & G141 & 2459005.73\\
\hline
\multicolumn{4}{c}{Inter-visit gap: 744 days + 894 min}\\
\hline
\multicolumn{4}{c}{Visit 3}\\
\hline
8 & Imaging & F127M & 2459750.38\\
  & Spectroscopic & G141 & 2459750.39\\
9 & Imaging & F127M & 2459750.52\\
  & Spectroscopic & G141 & 2459750.52\\
10 & Imaging & F127M & 2459750.58\\
   & Spectroscopic & G141 & 2459750.58\\
11 & Imaging & F127M & 2459750.65\\
   & Spectroscopic & G141 & 2459750.65\\
\enddata
%%%%%%%%%%%%%%%%%%%%%%%%%%%%%%%
\tablecomments{In each orbit, exposure times for Imaging observations were 59.2 seconds total (2 images $\times$ 29.6 seconds each) and Spectroscopic observations were 2504.8 seconds total (8 spectra $\times$ 313.1 seconds each). BJD$_{\rm{TDB}}$ here was calculated by converting from UT start date and time of observations (DATE-OBS and TIME-OBS in MJD in the \texttt{flt} fits headers) using an online applet developed by Jason Eastman \citep[]{Eastman10_UTC2BJD}.}
\end{deluxetable}

\begin{deluxetable}{crcr}
\tablecaption{Updated RECTE Parameters \label{tab:recte}}
\tablehead{
\colhead{Parameter} & \colhead{Value} & \colhead{Parameter} & \colhead{Value} 
}
%%%%%%%%%%%%%%%%%%%%%%%%%%%%%%%
\startdata
\hline
\multicolumn{4}{c}{Fixed Model Values}\\
\hline
$E_{\rm{f,tot}}$ &  225.7 & $E_{\rm{s,tot}}$ & 2192 \\
$\eta_{\rm{f}}$ & 0.0116 & $\eta_{\rm{s}}$ & 0.02075 \\
$\tau_{\rm{f}}$ & 3344 & $\tau_{\rm{s}}$ & 16,300 \\
\hline
\multicolumn{4}{c}{Best-fit MCMC Values}\\
\hline
$E_{\rm{f,0}}$(1) & 0.83 $\pm$ 0.05 & $E_{\rm{s,0}}$(1) & 0.15 $\pm$ 0.02 \\
$E_{\rm{f,0}}$(2) & 0.13 $\pm$ 0.06 & $E_{\rm{s,0}}$(2) & 0.23 $\pm$ 0.19 \\
$E_{\rm{f,0}}$(3) & 1.33 $\pm$ 0.09 & $E_{\rm{s,0}}$(3) & 0.72 $\pm$ 0.34 \\
\enddata
%%%%%%%%%%%%%%%%%%%%%%%%%%%%%%%
\tablecomments{Model parameters for the \texttt{RECTE} ramp-profile model, fast population on the left, slow population on the right. Globally fixed parameters for both fast and slow populations are shown in the top section. The bottom section shows parameters that were allowed to vary in the MCMC run. We treated the three visits separately, where Visits 1, 2 and 3 are denoted (1), (2), and (3) respectively.}
\end{deluxetable}

%%%%%%%%%%%%%%%%%%%%%%%%%%%%%%%%%%%%%%%%%%%%%%%%%%%%%%%%%%%%%%%%%%%%
%%%%%%%%%%%%%%%%%%%%%%%%%%%%%%%%%%%%%%%%%%%%%%%%%%%%%%%%%%%%%%%%%%%%
\section{Observations \& Data Reduction} \label{sec:obs_dataredux}
SDSS\,1557 was observed with Hubble Space Telescope (\textit{HST}) using the WFC3/IR/G141 grism setup as part of the HST program GO-15947 (PI: Apai). Ten orbits split equally between two visits were scheduled for 2020 June. During this run, three orbits failed owing to guide star acquisition failures, one during the first visit and two during the second visit. Due to the failures, a third visit (five additional orbits) was scheduled for 2022 June with the same instrument setup as visits 1 and 2. Four out of five orbits were successful during this third run, leading to a total of 11 successful orbits for SDSS\,1557 (summarized in Table~\ref{tab:obslog}).

At the beginning of each orbit, two direct images were obtained for wavelength calibration using the F127M filter, a 256×256 subarray setup, and the GRISM256 aperture. After the direct images, eight spectroscopic exposures of 313 s each were obtained using the G141 grism, a 256×256 subarray setup, and the GRISM256 aperture. This observing sequence was successfully carried out for 4 other WD--BD systems as part of the same HST program: SDSS~J141126.20+200911.1 \citep[]{Lew22}, WD 0137-349, EPIC 212235321 \citep[]{Zhou22}, and NLTT5306 \citep[]{Amaro23}.

The spectral extraction process is largely similar to that described in \citet{Amaro23}, based on an established pipeline that combines the WFC3/IR spectroscopic software \texttt{axe} \citep[]{Kummel09_aXe} and custom \texttt{Python} scripts. This pipeline has successfully extracted time-resolved observations of BDs \citep[]{Buenzli12, Apai13} and WD--BD binaries \citep[]{Zhou22, Lew22, Amaro23}.

\begin{figure*}
\begin{center}
\includegraphics[width=0.75\textwidth]{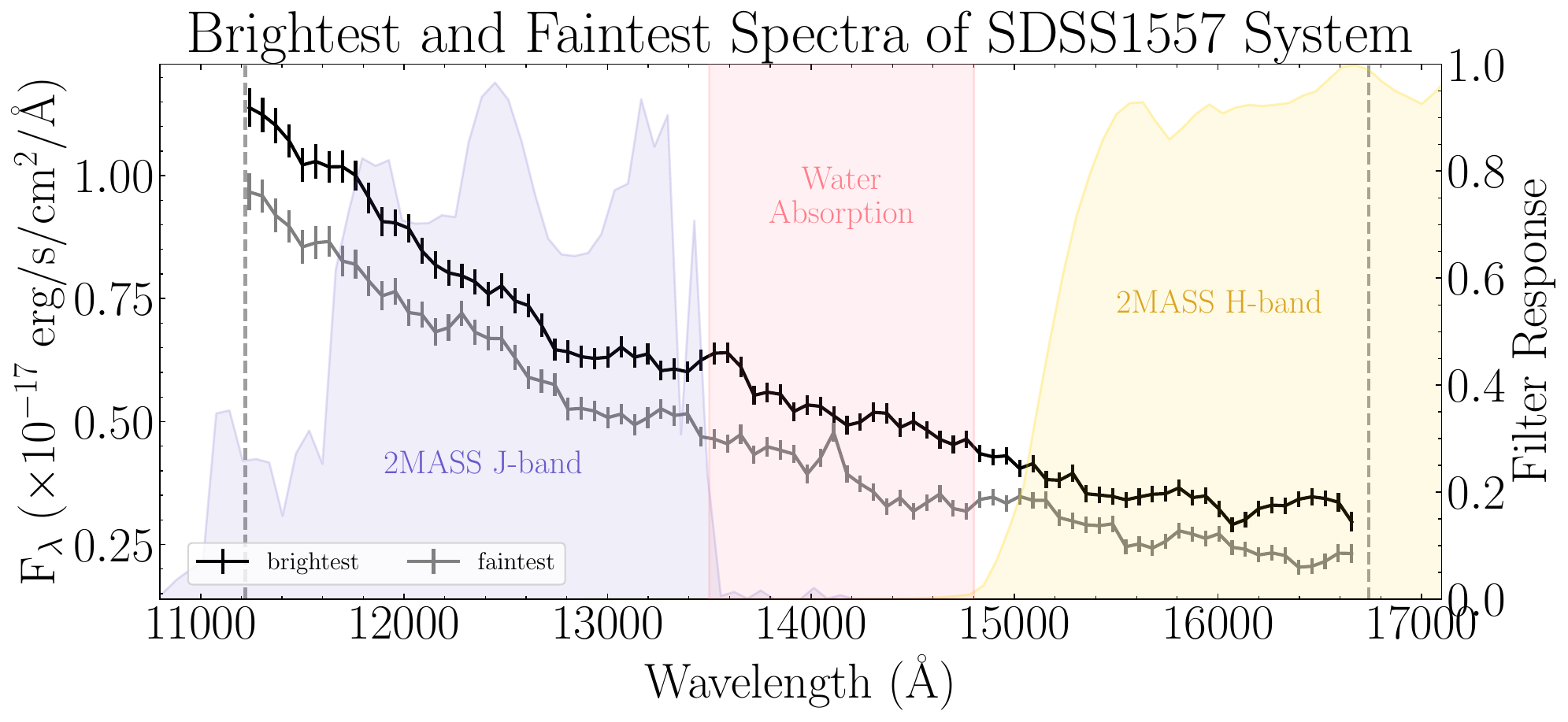}
\caption{Brightest and faintest spectra of SDSS\,1557 with wavelength cutoffs (vertical dashed lines; described in Section 4). Filters profiles, shown in purple, pink, and yellow, were used to make sub--band light curves for better comparison with other observations. The filter responses were re-binned to the resolution of our data and truncated according to our wavelength cutoffs, which led to a modified version of the 2MASS $H$--band: $H'$--band.}
\label{fig:science_spec_withfilters}
\end{center}
\end{figure*}

We downloaded the CalWFC3 (version 3.5.2) pipeline product \texttt{flt} files for SDSS\,1557 from the Barbara A. Mikulski Archive for Space Telescopes (MAST)\footnote{\url{https://archive.stsci.edu/index.html}}. We ensured precise wavelength calibration for each orbit by using Source Extractor \citep[]{Bertin96_sextractor} on the median combined direct images from the beginning of each orbit. Then we took special care to manually check the positions of bad pixel flags in the G141 observations if they were near the grism spectrum. The same care was taken when checking the master sky image \texttt{WFC3.IR.G141.sky.V1.0.fits}. Configuration and reference files were downloaded from the \texttt{aXe} (\texttt{hstaxe}) WFC3 Extraction Example Cookbook\footnote{\url{https://github.com/npirzkal/aXe_WFC3_Cookbook}}. We then extracted the 1D spectra using a four pixel radius extraction window, resulting in a spectral resolution of $\approx$27 near 1.4 \mum{}.

Next, we corrected the \textit{HST} ramp effect --- caused by charge trapping and delayed release --- by fitting a \texttt{RECTE} model \citep[]{Zhou17} to the broadband light curve. The process for fitting the \texttt{RECTE} model is described in further detail in \citet{Amaro23}. To summarize, we use a custom Markov Chain Monte Carlo (MCMC) performed by \texttt{emcee} \citep[]{emcee} to fit 6 \texttt{RECTE} parameters, two per visit. The results of the MCMC are presented in Table~\ref{tab:recte} and the final ramp model made with these results was divided out of each spectrum.

We then resampled our spectra to the resolving power of the observations using the 2D stamps output \texttt{axecore} from \texttt{hstaxe}. We plotted the counts of each pixel in the direction perpendicular to the dispersion axis and fit the shape of the count curves with a Moffat profile. The FWHM values were converted from pixel space to wavelength space, resulting in a spectral resolution of R$\sim$215 at 14000 \AA{}. The errors on the resampled spectra were calculated according to a method from \citet{Carnall17} (see Equation 1 in \citealt{Amaro23}).

Finally, to remove unreliable data points, we performed wavelength cutoffs wherever the resampled spectral errors were greater than two times the average error value between 13,00 and 15000 \AA{}. The resulting science spectra of SDSS\,1557 is shown in Figure~\ref{fig:science_spec_withfilters}.

%%%%%%%%%%%%%%%%%%%%%%%%%%%%%%%%%%%%%%%%%%%%%%%%%%%%%%%%%%%%%%%%%%%%
%%%%%%%%%%%%%%%%%%%%%%%%%%%%%%%%%%%%%%%%%%%%%%%%%%%%%%%%%%%%%%%%%%%%
\section{Light Curve Analysis} \label{sec:lcanalysis}

\subsection{Broadband Light Curve}
\label{sec:bb_lc}

Studying the light curves of systems with tidally-locked companions reveals the presence, or absence, of any longitudinal intensity modulations that may be present within the photosphere.  In the case of ultra-short period WD--BD binaries, we assume constant flux from the WD primary, with any variation on the surface being negligible compared to the BD companion. To begin characterization of the flux modulations coming from the rotation of the BD companion, we created simple phase curve models that should emulate most of the periodic behavior.

\begin{figure*}
\begin{center}
\includegraphics[width=\textwidth]{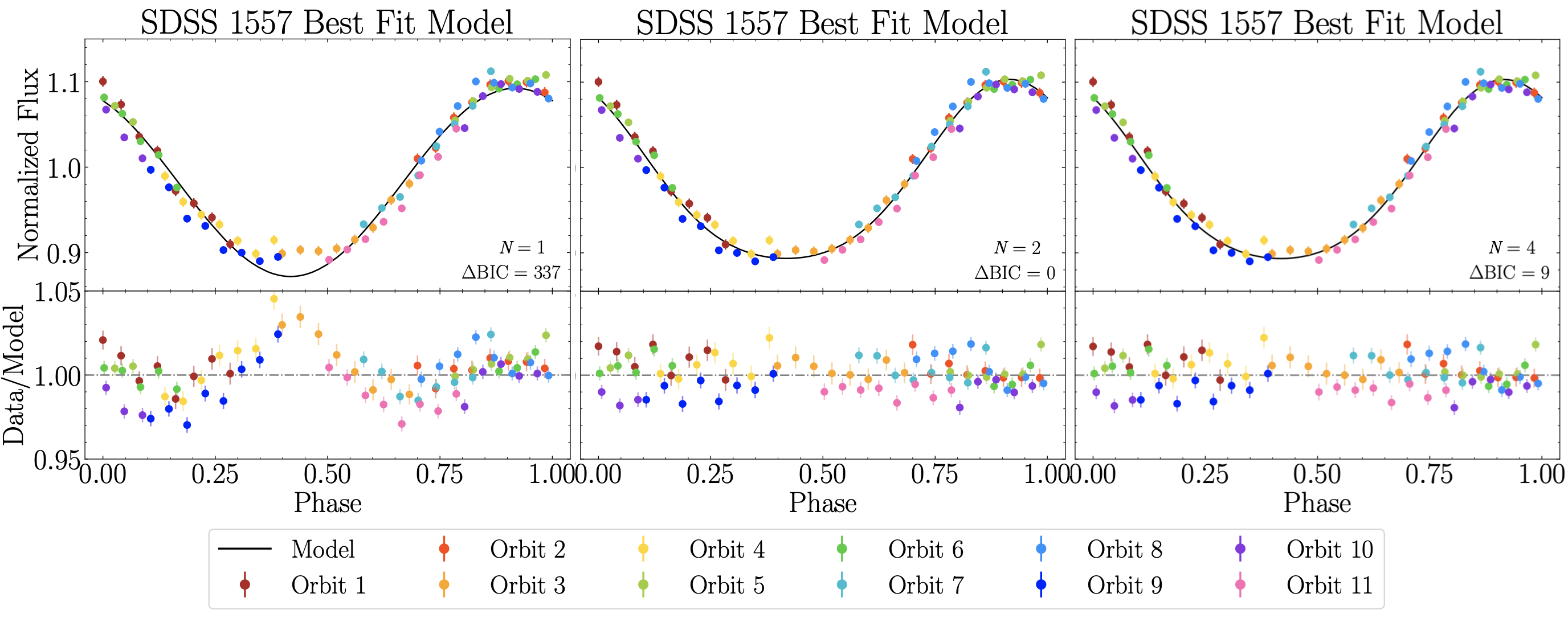}
\caption{Comparison of best fitting Fourier Series models (solid black lines) for orders $N$=1, 2, and 4. Each of the 11 HST orbits are represented in a different color. Based on BIC values, the $N$=2 model is marginally favored for SDSS\,1557, with a model that captures light curve behavior to within 2.5\%.}
\label{fig:pcmodels_perorbit}
\end{center}
\end{figure*} 

Once the \texttt{RECTE} ramp models were divided out and each spectrum was resampled, we determined the best-fit phase curve model to the broadband light curve with an MCMC. Similar to \citet{Amaro23}, the number of phase curve parameters varied according to the number of sine curves in each model. Specifically, we considered combinations of Fourier series from:
\begin{equation}
    \label{eq:pcmodel_fourier}
    F(t) = F_{\rm{norm}} + \sum_{k} a_{k} \sin (k \frac{2 \pi t}{P}) + 
    b_{k} \cos (k \frac{2 \pi t}{P}),
\end{equation}
where $F(t)$ is the phase curve as a function of observation time, $F_{\rm{norm}}$ is the scaling parameter for normalization (always near 1.0), $P$ is the period (fixed to published value from \citealt{Farihi17}), $a_k$ and $b_k$ are the amplitudes for sine and cosine, and $k$ is the order of the sinusoidal function. Amplitudes were calculated following Equation (2) in \citet{Zhou22}, defined as: amp$_{k} = \sqrt{a_{k}^{2}+b_{k}^{2}}$. We considered three combinations of $k$=1, 2, and 4 (i.e. $N$=1 is only $k$=1; $N$=2 is $k$=1 and 2; and $N$=4 is $k$=1, 2, and 4). The best fitting parameters and Bayesian information criterion (BIC) values are presented in Table~\ref{tab:pc_params}.

Once models were created, we period-folded the data and calibrated phase = 0 to the time of the first observation (Table~\ref{tab:obslog}). A comparison of each period-folded phase curve for $N=$ 1, 2, and 4 are shown in Figure~\ref{fig:pcmodels_perorbit}, where the $N$=2 model was the preferred fit according to the BIC values. All best-fit phase curve model parameters and corresponding BIC values for the broadband light curve are presented in the top left section of Table~\ref{tab:pc_params}.

In the $N$=2 broadband phase curve of SDSS\,1557, the amplitudes were $amp_1$ = 0.105 $\pm$ 0.001 and $amp_2$ = 0.015 $\pm$ 0.001. We observed a narrow peak and a wider base. We believe this suggests a surface flux distribution with a compact, extremely hot substellar area, caused by the tidally-locked irradiation, and a broader area with lower temperature, caused by a low heat redistribution fraction. 

An ephemeris for the orbit of SDSS\,1557 was published in \citet{Farihi17}, where the authors use radial velocities from  VLT+XSHOOTER observations from 2014\footnote{P.I.: S. Parsons; Program I.D. 097.C$-$0386} and 2016\footnote{P.I.: J. Farihi; Program I.D. 093.D$-$0030}. A precise ephemeris can be used to determine the presence of an absolute phase shift in the substellar hot spot, which would suggest the presence of winds strong enough to shift the hottest point in the atmosphere. However, when we extrapolate the published ephemeris to the date of our HST observations, the uncertainty grows to be of the same order of magnitude as our uncertainty in phase (0.02 phase). For future atmospheric observations of this binary system, we recommend observing precise radial velocities at or near the time of observation in order to quantify any hot spot shift.

%%%%%%%%%%%%%%%%%%%%%%%%%%%%%%%%%%%%%%%%%%%%
%%%%%%%%%%%%%%%%%%%%%%%%%%%%%%%%%%%%%%%%%%%%
\subsection{Sub--band Light Curves}
\label{sec:subband_lc}

In addition to analyzing broadband data, studying light curves generated from sub--bands offers a valuable opportunity to investigate latitudinal brightness variations as a function of atmospheric pressure. This is because different wavelength ranges probe different atmospheric pressure levels. In brown dwarf studies, it is common to use $J$-- and $H$--band light curve filters \citep[e.g.][]{Buenzli12}. Thus, we derived $J$-- and $H'$--band light curves from our spectral observations. To create the sub--band $J$--band light curve, we convolved each spectrum with the Two-Micron All Sky Survey (2MASS) $J$--band transmission filter from the 2MASS All-Sky Data Release \citep{2MASS_Skrutskie06} from 11000 \AA{} to 13500 \AA{}. We used the same method to create the $H'$--band light curve, except only we integrated from 15000 \AA{} to the end of each spectrum (approximately 16700 \AA{}), since the 2MASS $H$--band transmission filter extended beyond our spectra. In addition, we created a third sub--band light curve in the water--band by integrating from 13600 to 14800 \AA{}. Figure~\ref{fig:science_spec_withfilters} shows each filter response function and their respective wavelength ranges.

Real atmospheres are complex, and different wavelengths probe different pressures due to strongly wavelength-dependent gas and dust opacities \citep[e.g.][]{Buenzli12, Yang16}. Thus, by comparing light curves from different filters, we can directly compare brightness distributions in different atmospheric layers. Similarities or differences between the light curves allow us to identify pressure-dependent structure and dynamics within the atmosphere. To quantify any similarities or differences between our three sub--band light curves (i.e., $J$--, water--, and $H'$--band), we created phase curve models using the same approach described in Section~\ref{sec:bb_lc}.

\begin{figure}
\begin{center}
\includegraphics[width=0.48\textwidth]{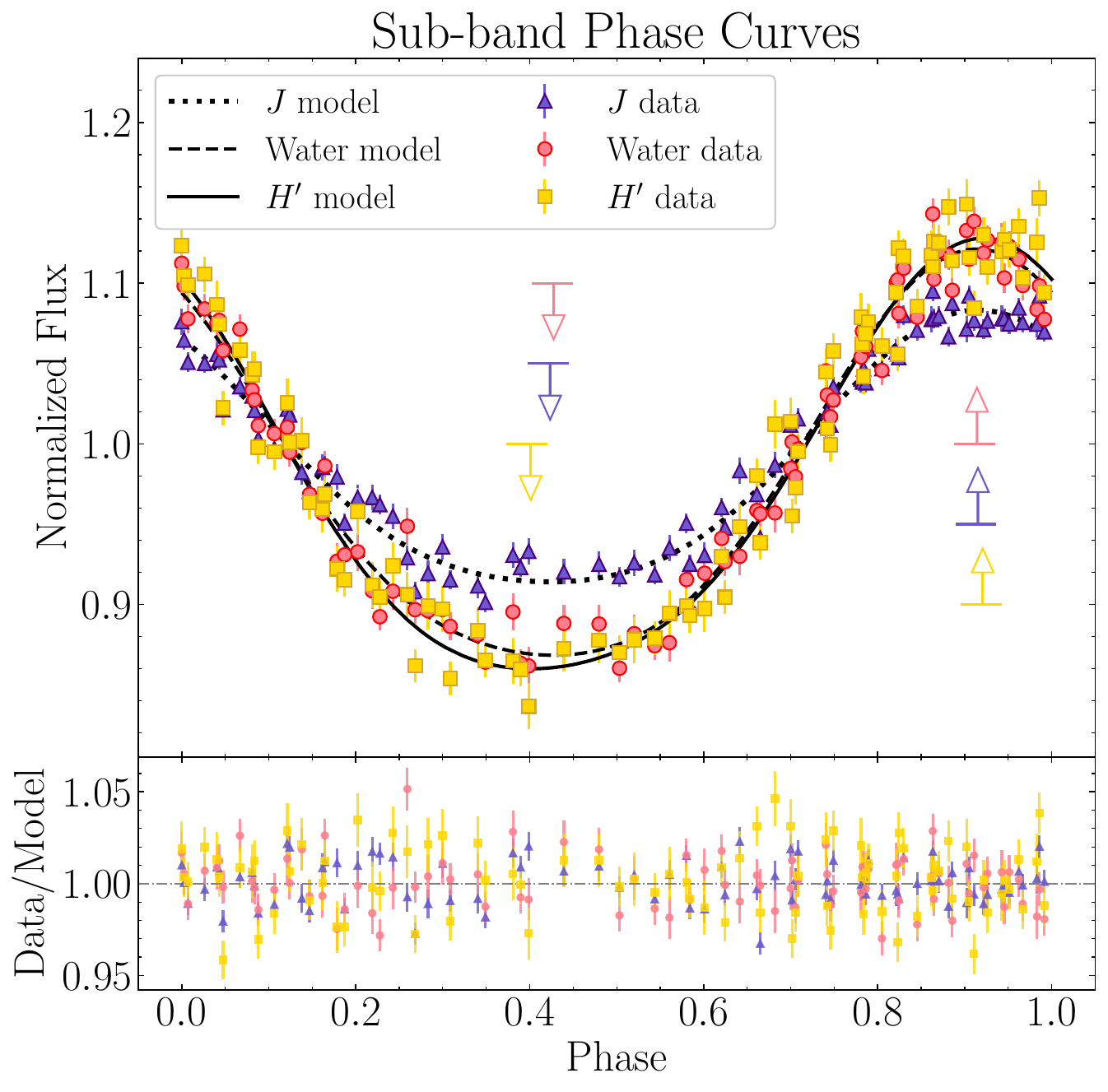}
\caption{Sub--band phase curves made in the 2MASS $J$-- (indigo triangles), Water-- (coral circles), and 2MASS $H'$--bands (yellow squares). Best-fit models for each filter are presented as well as precise phases for each peak and valley (phase uncertainty shown as horizontal bars on arrows). We observe no significant relative phase shift between the peaks or troughs of the phase curves. The Water-- and $H'$--band phase curves both exhibit amplitudes greater than 12\%, whereas the $J$--band phase curve exhibits a shallower amplitude around 8\%.}
\label{fig:subband_pc}
\end{center}
\end{figure}

The best-fit parameters from all three models ($N$ = 1, 2, and 4) for each sub--band light curve are presented in Table~\ref{tab:pc_params}. Similar to the broadband light curve, the $N$=2 model yielded the best-fit. The best-fit $amp_1$ amplitudes for the Water-- and $H'$--band phase curve models were 0.126 $\pm$ 0.002 and 0.133 $\pm$ 0.002, respectively. Interestingly, the $J$--band model had a significantly shallower amplitude of 0.084 $\pm$ 0.001. 
% J (amp1) = 0.084504 +- 0.001; Water (amp1) = 0.126372 +- 0.002; H' (amp1) = 0.133544 +- 0.002
We see no evidence for relative phase shifts between the peaks of the sub--band light curves: $\phi_{\rm{peak}}$($J$) $-$ $\phi_{\rm{peak}}$(Water) = 0.001 $\pm$ 0.021; $\phi_{\rm{peak}}$($J$) $-$ $\phi_{\rm{peak}}$($H'$) = 0.005 $\pm$ 0.021. These phase shifts are relative to the start of the first observations, which is arbitrary. We are therefore interested in the difference between these phase values. Light curves and best-fit models are shown in Figure~\ref{fig:subband_pc}.

\subsection{Spectroscopically-resolved Phase Curves}
\label{sec:amp_wav}
The shallower amplitude observed in the sub--band $J$--band phase curve suggested a potential wavelength dependence. To explore this possibility, we applied more precise wavelength bins in our analysis of the phase curve models derived from the spectra. The narrower wavelength bins had a width of $\approx$447 \AA{} and offered an average signal-to-noise ratio (S/N) of 80. Building upon the success of the $N$=2 model for broadband and sub--band light curves, we focused solely on the $N$=2 model for these spectroscopically-resolved phase curves.

\begin{figure}
\begin{center}
\includegraphics[width=0.48\textwidth]{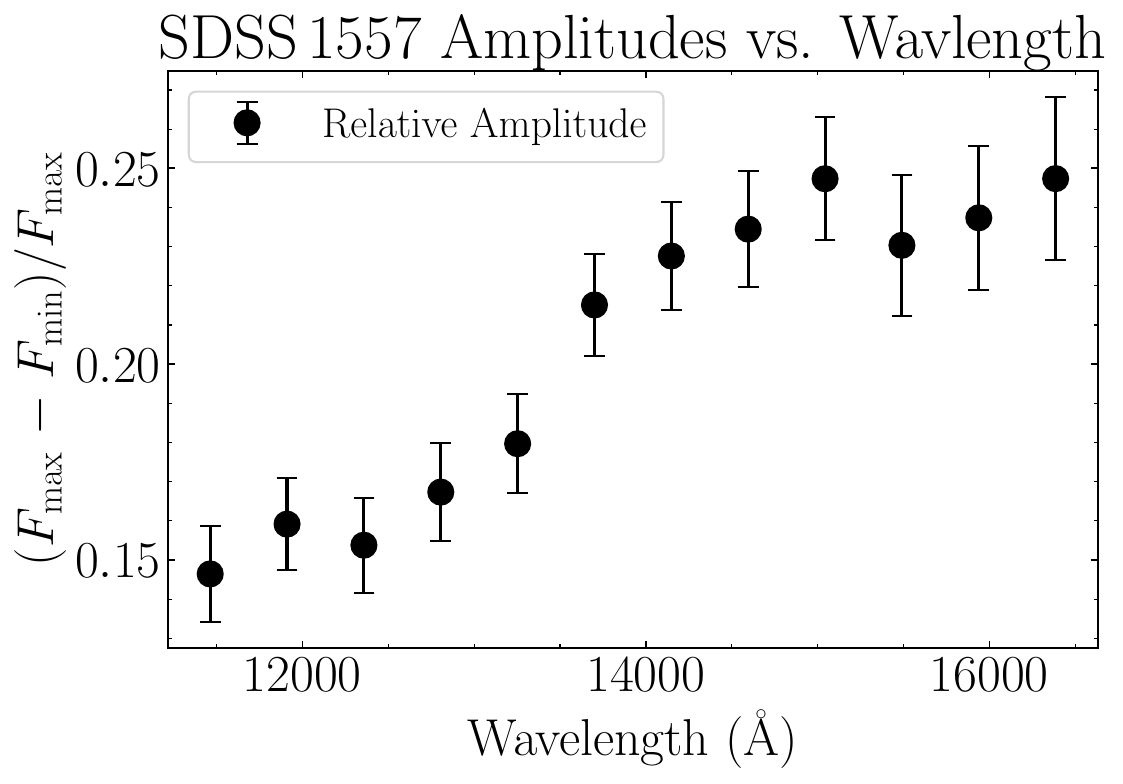}
\caption{Relative phase curve amplitudes as a function of wavelength for SDSS\,1557. In this case, relative refers to the total change in flux relative to the maximum flux. Overall, the amplitude changes tend to increase with increasing wavelength.}
\label{fig:amps_wave}
\end{center}
\end{figure}

In Figure~\ref{fig:amps_wave}, the best-fitting amplitudes, shown as $\Delta F$/$F_{\rm{max}}$, are presented as a function of wavelength. Within the range of 1.14 to 1.50 \mum{}, the relative amplitude differences increased from $\sim$15 to $\sim$25 percent, with the greatest increase occurring between 1.32 and 1.44~$\mu$m. Past 1.50~$\mu$m, within the $H'$--band, the differences level-off, hovering around an average of 23.5 percent.

Spectroscopically-resolved phase curves and the analysis of wavelength dependence on their parameters, provides us a glimpse into the longitudinal-vertical circulation patterns within an atmosphere, as wavelength can be used a tracer for atmospheric pressure \citep[e.g.,][]{Yang16}. By anchoring our interpretations using these methods, we can better understand the dynamics at play. In Figure~\ref{fig:amps_wave}, we observe that longer wavelengths generally exhibit stronger modulations. This correlation can be attributed to the BD, which is believed to be the primary contributor to the emission flux at longer wavelengths, due to relative temperature differences.

\begin{deluxetable*}{lccc|lccc}
\tablecaption{Best-fit Phase Curve Parameters\label{tab:pc_params}}
\tablehead{
\colhead{Parameter} & \colhead{$k$=1} & \colhead{$k$=2} &  \colhead{$k$=4} & \colhead{Parameter} & \colhead{$k$=1} & \colhead{$k$=2} &  \colhead{$k$=4} }
%%%%%%%%%%%%%%%%%%%%%%%%%%%%%%%
\startdata
\hline
\multicolumn{4}{c|}{Broadband Phase Curve} & \multicolumn{4}{c}{$J$--band Phase Curve}\\
\hline
$F_{\rm{norm}}$ &  0.982 $\pm$ 0.001 & 0.984 $\pm$ 0.001 &  0.984 $\pm$ 0.001 & $F_{\rm{norm}}$ & 0.987 $\pm$ 0.001 & 0.986 $\pm$ 0.001 & 0.987 $\pm$ 0.001 \\
$a_1$ & -0.084 $\pm$ 0.001 &  0.067 $\pm$ 0.001 &  0.067 $\pm$ 0.001 & $a_1$ & -0.068 $\pm$ 0.001 &  0.073 $\pm$ 0.001 &  0.073 $\pm$ 0.001 \\
$b_1$ &  0.071 $\pm$ 0.001 & -0.081 $\pm$ 0.001 & -0.081 $\pm$ 0.001 & $b_1$ &  0.057 $\pm$ 0.001 & -0.042 $\pm$ 0.001 & -0.042 $\pm$ 0.001 \\
$a_2$ & ...                & -0.003 $\pm$ 0.001 & -0.003 $\pm$ 0.001 & $a_2$ & ...                &  0.006 $\pm$ 0.001 &  0.006 $\pm$ 0.001 \\
$b_2$ & ...                & -0.015 $\pm$ 0.001 & -0.015 $\pm$ 0.001 & $b_2$ & ...                & -0.011 $\pm$ 0.001 & -0.011 $\pm$ 0.001 \\
$a_4$ & ...                & ...                &  0.000 $\pm$ 0.001 & $a_4$ & ...                & ...                &  0.000 $\pm$ 0.001 \\
$b_4$ & ...                & ...                & -0.000 $\pm$ 0.001 & $b_4$ & ...                & ...                & -0.002 $\pm$ 0.001 \\
BIC & -80.892 & -418.270 & -409.353 & BIC & -312.2030 & -425.111 & -420.429 \\
\hline
\multicolumn{4}{c|}{Water--band Phase Curve} & \multicolumn{4}{c}{$H'$--band Phase Curve}\\
\hline
$F_{\rm{norm}}$ & 0.975 $\pm$ 0.001 &  0.977 $\pm$ 0.001 &  0.977 $\pm$ 0.001 & $F_{\rm{norm}}$ &  0.972 $\pm$ 0.001 &  0.974 $\pm$ 0.001 &  0.974 $\pm$ 0.001\\
$a_1$ & -0.102 $\pm$ 0.002 &  0.110 $\pm$ 0.002 &  0.110 $\pm$ 0.002 & $a_1$ & -0.108 $\pm$ 0.002 &  0.116 $\pm$ 0.002 &  0.116 $\pm$ 0.002\\
$b_1$ &  0.085 $\pm$ 0.002 & -0.063 $\pm$ 0.002 & -0.063 $\pm$ 0.002 & $b_1$ &  0.091 $\pm$ 0.002 & -0.067 $\pm$ 0.002 & -0.067 $\pm$ 0.002\\
$a_2$ & ...                &  0.008 $\pm$ 0.002 &  0.007 $\pm$ 0.002 & $a_2$ & ...                &  0.012 $\pm$ 0.002 &  0.012 $\pm$ 0.002\\
$b_2$ & ...                & -0.016 $\pm$ 0.002 & -0.016 $\pm$ 0.002 & $b_2$ & ...                & -0.016 $\pm$ 0.002 & -0.016 $\pm$ 0.002\\
$a_4$ & ...                & ...                & -0.002 $\pm$ 0.001 & $a_4$ & ...                & ...                &  0.002 $\pm$ 0.002\\
$b_4$ & ...                & ...                &  0.002 $\pm$ 0.002 & $b_4$ & ...                & ...                &  0.000 $\pm$ 0.002\\
BIC & -325.548 & -449.537 & -443.056 & BIC & -276.441 & -367.102 & -359.030\\
\enddata
%%%%%%%%%%%%%%%%%%%%%%%%%%%%%%%
\tablecomments{The $k$=2 phase curve was always best-fit according to the BIC values.}
\end{deluxetable*}

%%%%%%%%%%%%%%%%%%%%%%%%%%%%%%%%%%%%%%%%%%%%%%%%%%%%%%%%%%%%%%%%%%%%
%%%%%%%%%%%%%%%%%%%%%%%%%%%%%%%%%%%%%%%%%%%%%%%%%%%%%%%%%%%%%%%%%%%%
\vspace{-0.5 cm}
\section{Rotational Phase-Resolved Spectra} \label{sec:bdspec}

The $\sim$1.1 to 1.7 $\mu$m wavelength coverage of \textit{HST}/WFC3/G141 is excellent for constraining the atmospheric properties of irradiated brown dwarfs, since their flux contributions peak in the NIR. Further aided by the favorable flux contrast ($\sim$10--30\%) between the irradiated BD companion and WD primary, in NIR wavelengths, we are able to spectrally and spatially map irradiated BD atmospheres in great detail. Thus, we are also able to facilitate comparisons to current models of hot Jupiter and BD atmospheres.

\subsection{Modeling the Debris Disk Contribution}
Among our sample of six WD--BD systems, SDSS\,1557 is unique due to a third component in the system, its circumbinary debris disk. \citet{Farihi17} initially identified this disk and calculated that it should exist at an orbital radius of 3.3 $R_{\odot}$ at a temperature of $T$$\approx$1100~K. We modeled the expected flux contribution of the disk at our \textit{HST} wavelengths, and determined that the disk excess emission at these wavelength is negligible compared to the brightness of the WD--BD binary, with $F_{\rm{disk}}$/$F_{\rm{WD-BD}}$ ranging from 0.04 to 1.3\%. Therefore, our analysis cannot yield significant insights into SDSS\,1557's debris disk. For a comprehensive analysis on the disk structure, evolution, and role within the SDSS\,1557 system, we refer to \citet{Farihi17} and \citet{Swan20}. We note, that upcoming studies with the James Webb Space Telescope --- with its far greater sensitivity at longer wavelengths --- will likely provide powerful constraints on the disk composition and structure.

\subsection{Modeling the White Dwarf Contribution} \label{sec:wdspec}

To study the atmosphere of the irradiated BD in more detail, we modeled and subtracted the WD contribution from our phase-resolved spectra. We began the modeling process by downloading a model grid for pure-hydrogen atmosphere WDs from \citet{Koester10}, which features WD atmosphere models based on two parameters: $T_{\rm{eff}}$ (effective temperature) and log($g$) (surface gravity). We bilinearly interpolated along the grid to the published values for SDSS\,1557 (see Table~\ref{tab:keyprops}). Uncertainty in the adopted values was accounted for by creating 10,000 interpolated WD models, with $T_{\rm{eff}}$ and log($g$) values sampled from Gaussian distributions. For the Gaussian distributions, the means were set to 21800 K and 7.63 cm s$^{-2}$, with corresponding standard deviations of 800 K and 0.11 cm s$^{-2}$. To quantify the uncertainty, we measured the Full-Width-Half-Maximum (FWHM) of the histogram of flux values at each wavelength. These model uncertainties are represented as solid black error bars in Figure~\ref{fig:wd_model}.

Next, we needed to scale the flux of our WD model to reproduce the expected flux at the distance and radius of our WD. The previously published value for distance was 520$\pm$35 pc \citep{Farihi17}. Using more recent, photogeometric data from \textit{Gaia} Data Release 3, we updated values for the radius and distance to $R_{\rm{WD}}$ = 0.0162 $\pm$ 0.0012 R$_{\odot}$ and $D$ = 500$^{+19.8}_{-18.0}$ pc (refer to Appendix~\ref{sec:updateparams} for further details on this calculation). The updated distance is within the 1$\sigma$ uncertainty range of the previous value. We then multiplied our WD model by the scale factor $\big( R_{\rm{WD}}^2 / D^2 \big)$ and propagated uncertainties associated with the WD's radius and distance. Consequently, there are two sources of uncertainty in modeling the WD contribution: the first arising during the model interpolation and the second introduced during the subsequent flux scaling. In Figure~\ref{fig:wd_model}, the total uncertainty, considering both uncertainty sources, is represented by the gray shaded region.

\subsection{Extracting Phase-resolved\\Spectra of SDSS\,1557B}

In order to understand the impact of external irradiation on an ultracool atmosphere, we set out to isolate the BD spectra from our observations of the non-spatially resolved system. To start, we focused on day vs. night hemispheres, as these phases should show the greatest differences in a tidally-locked system. Previous studies of ultra-short period WD--BD systems have shown brightness temperature differences between day and night hemispheres, which could be an indicator for differences in cloud coverage, atmospheric composition, or a combination of both.

For our SDSS\,1557 data, we applied a wavelength shift to each spectrum to account for any radial velocity shifts before subtracting the WD model. We determined the magnitude of the radial velocity shift ($v_r$) at a certain phase using the following equation:
\begin{equation}
    v_r = \gamma_1 + K_1 \rm{sin}(2 \pi \phi) - \big(\gamma_2 + K_2 \rm{sin}(2 \pi \phi) \big).
\end{equation}
Here, the phase ($\phi$) of each spectrum was calibrated to the ephemeris and period from \citet{Farihi17}. The parameters $\gamma_1$ and $K_1$ represent the WD's radial velocity, while $\gamma_2$ and $K_2$ correspond to the brown dwarf's contribution. We adopted the values for $\gamma_1$, $K_1$, $\gamma_2$, and $K_2$ from \citet{Farihi17} as well.

Next, we performed a shift on each spectrum to its rest wavelength using the Doppler shift equation:
\begin{equation} \label{eq:doppshift}
\frac{v_r}{c} = \frac{\lambda_{\rm{obs}} - \lambda_{\rm{rest}}}{\lambda_{\rm{rest}}}
\end{equation}
Solving for $\lambda_{\rm{rest}}$ in Equation~\ref{eq:doppshift}, the largest wavelength shift observed in our sample was less than 20 \AA{}; i.e. less than a third of our spectral resolution of 65.27 \AA{}.

\begin{figure*}
\begin{center}
\includegraphics[width=0.78\textwidth]{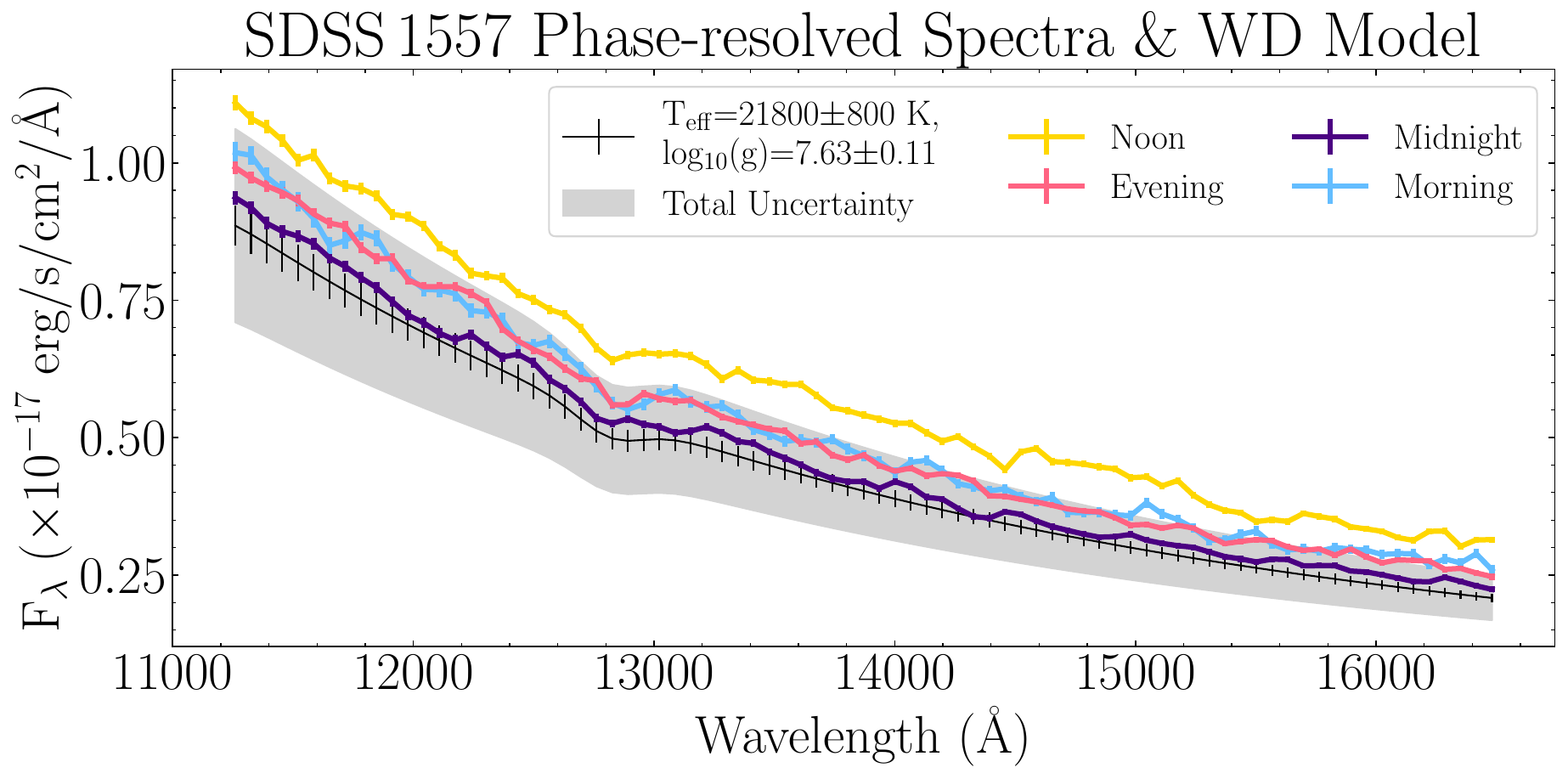}
\caption{The modeled flux contribution of the WD SDSS\,1557A, shown with both sources of uncertainty. The first being in the creation of the model (black errorbars) and the second being in the flux scaling (gray shaded region). Additionally, the phase-resolved spectra of the irradiated BD SDSS\,1557B are shown at 4 key flux phases: Noon, Evening, Midnight, and Morning.}
\label{fig:wd_model}
\end{center}
\end{figure*}

To ensure data quality and minimize spectral variations, we derived the BD spectra at representative phases by calculating the median of five spectra. For ``Noon" and ``Midnight", we used five spectra that corresponded to the five brightest and faintest broadband light curve points. For ``Morning" and `Evening", we identified the appropriate phase by adding and subtracting 0.25 from the phase with the lowest absolute flux in the broad band light curve. Then, we found the five spectra nearest to these new phases and took their median. The four phase-resolved spectra are presented in Figure~\ref{fig:wd_model}.

\begin{figure}
\begin{center}
\includegraphics[width=0.47\textwidth]{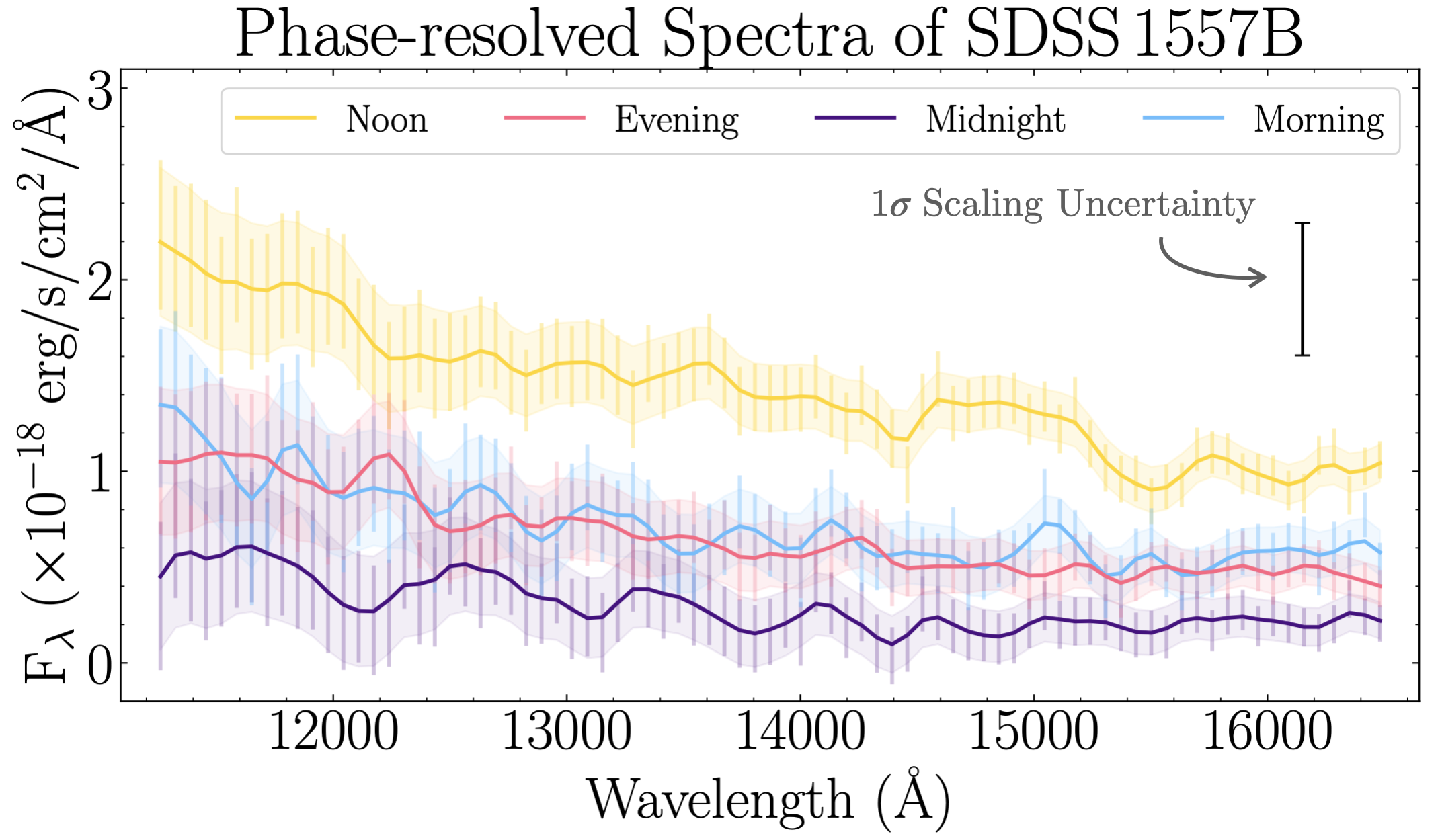}
\caption{Derived phase-resolved NIR spectra of irradiated BD SDSS\,1557B. Vertical errorbars represent systematic uncertainty originating from the creation of the WD model. The size of scaling uncertainties for both hemispheres is labeled near the upper right. Solid lines and shaded regions depict our observations and uncertainties convolved with a Gaussian kernel with a width equal to the spectral resolution of the G141 grism.}
\label{fig:bd_spec}
\end{center}
\end{figure}

We then proceeded to subtract the WD model and isolate the irradiated BD components, shown in Figure~\ref{fig:bd_spec}. The noon spectrum presents as a nearly featureless slope, likely due to the intense irradiation received by the WD primary. The midnight spectrum becomes nearly undetectable with flux levels near zero and S/Ns below 3$\sigma$ for all wavelengths. The morning and evening spectra exhibit no significant features, but are both nearly half the flux of the noon spectrum.

\subsection{Day and Night Brightness\\Temperatures of SDSS\,1557B} \label{brighttemps}

The thermal structure in the atmosphere of a stellar companion is influenced by various factors, including internal heat flux, the profile of net absorbed stellar flux, and opacity structure \citep[][]{MarleyRobinson15}. To explore the atmospheric pressure-temperature structure, we determined brightness temperatures ($T_B$) as a function of wavelength, i.e.,  the temperatures at which a blackbody of the brown dwarf's size would emit the same specific intensity as the observed flux values. 
\begin{equation}
    T_B = \frac{hc}{\lambda k} \rm{ln} \Big( \frac{2 \pi h c^2 R_{BD}^2}{F_{BD} D^2 \lambda^5} + 1 \Big)^{-1}
\end{equation}
Expressing emission spectra as $T_B$ is widely used in BD and exoplanet studies, as well as previous works on our targets \citep[][]{Casewell15, Casewell18a, Lew22, Zhou22, Amaro23}. By combining brightness temperatures with a Pressure-Temperature (P-T) profile, we can effectively map the relative pressure regions associated with each wavelength, revealing the vertical structure of the atmosphere during each observed phase \citep[][]{Yang16}.

\begin{figure}
\begin{center}
\includegraphics[width=0.48\textwidth]{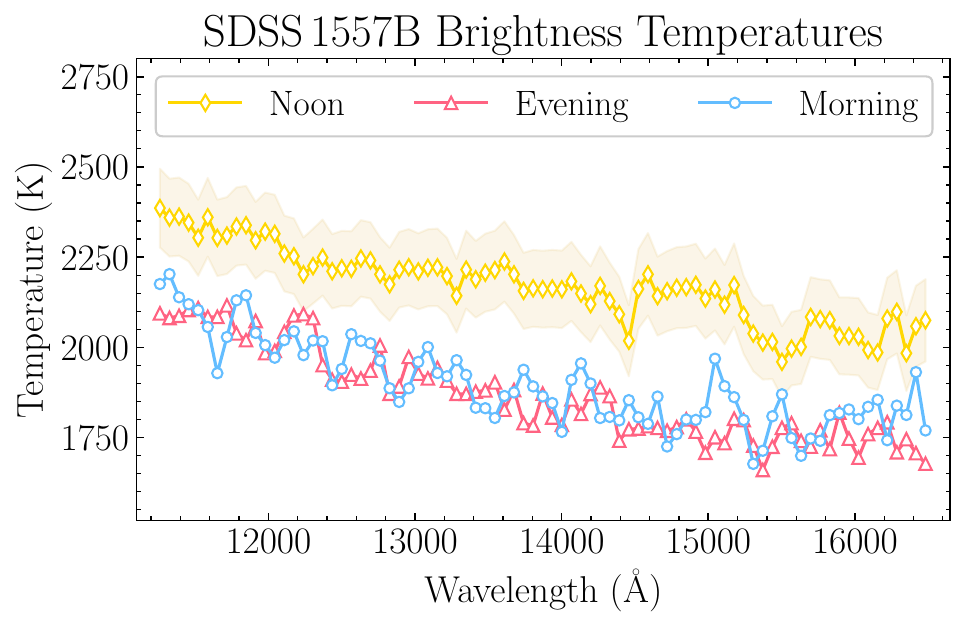}
\caption{Brightness temperatures of SDSS\,1557B, calculated by solving the Planck equation for $T_{\rm{B}}$ at each wavelength. Errorbars for the noon spectrum are shown as the shaded region, whereas errors for evening and morning are excluded for easier viewing. Errors for evening and morning are 95~K average.}
\label{fig:brighttemps}
\end{center}
\end{figure}

Figure~\ref{fig:brighttemps} presents the calculated brightness temperatures versus wavelength for the four extracted spectra of SDSS\,1557B. In the noon spectrum of SDSS\,1557B, $T_B$ decreases nearly monotonically towards longer wavelengths, from 2600 K at 1.12 \mum{} to 2250 K at 1.56 \mum{}. We do observe two increases in $T_B$ that deviate from the monotonic decline: the first centered around 1.5 \mum{} and the second starting at 1.56 \mum{} until 1.67 \mum{}. The morning and evening spectra exhibit similar slopes as the noon spectrum, and are qualitatively the same except for deviations at 15000 \AA{} and 16500 \AA{}.

The night side of SDSS\,1557B is particularly faint and did not result in confident detections: At its original resolution of approximately 65 \AA{} per data point, the spectrum primarily consisted of non-detections. We calculated upper limits for brightness temperatures by binning the spectra in wavelength, thus increasing the signal-to-noise ratio (S/N) for each bin. A S/N value greater than or equal to 3 was considered a detection, while values greater than or equal to 2 were suitable for establishing upper limits. We gradually increased the bin size in the spectrum and ultimately settled on a spectrum with a resolution of 965 \AA{} per data point, as this was the only binning option that yielded more than one data point with an S/N of 2 or higher. The upper limits for those two data points are shown in Figure~\ref{fig:brighttemps}. 

Given the temperature estimate, the day side is likely too hot to host silicate clouds \citep{Lunine86, Lunine89, BurrowsSharp99, Marley02}, but H$^-$ opacity may contribute to the spectral slope \citep[][]{Arcangeli18}. This scenario would argue for the absence of solid-state silicate absorption on the day side and a temperature inversion. However, in the morning and evening longitudes, the Si--O solid state (stretching--bending) absorption feature may be present -- testing this will require observations at longer wavelengths, near 10 \mum{} \citep[e.g.,][]{Apai2005,Luna2021,Suarez22}. A known signature of temperature inversion is the Paschen series atomic hydrogen n=5$\rightarrow$3 emission feature at 1.28 \mum{}. This signature may be present in the day.

Based on the upper limits for midnight brightness temperatures, temperatures exceeding 1500~K between 12000 and 14000 \AA{} are not anticipated. If we assume the actual midnight brightness temperatures at a higher resolution exhibit a comparable pattern to those in other phases, it is reasonable to infer the presence of silicate cloud coverage in this hemisphere.

%%%%%%%%%%%%%%%%%%%%%%%%%%%%%%%%%%%%%%%%%%%%%%%%%%%%%%%%%%%%%%%%%%%%
%%%%%%%%%%%%%%%%%%%%%%%%%%%%%%%%%%%%%%%%%%%%%%%%%%%%%%%%%%%%%%%%%%%%
\section{Comparison to Brown Dwarf Models}
\label{sec:Compare_models}

\subsection{Irradiated Models}
\label{sec:Compare_irr_models}

We compared the phase-resolved spectra of SDSS\,1557B to a number of PHOENIX atmosphere model spectra. These models follow the same set-up as those of \cite{LothringerCasewell2020}. In short, the companion atmosphere is irradiated by a WD whose spectrum is modeled by the \citet{Koester10} grid, with fundamental parameters from \citep{Farihi17}. An iterative process is conducted to calculate the BD's thermal structure until it reaches radiative equilibrium. We run the model at a wavelength sampling of 1~\AA{} from 10~\AA{} to 5\mum{}, with coarser sampling out to 1000\,\mum{}, with the irradiation spectrum defined between 0.09 and 3\,\mum{}. The models used a rainout abundance scheme, where elements that participated in condensation lower in the atmosphere were depleted in the layers above (or at lower pressure than) the condensation level of that species.

We first started with a small grid of day side heat-redistribution models ($T_{irr}$ = 2900~K) at solar metallicity with a range of surface gravity (log(g)$_{cgs}$ = 4.95, 5.2, and 5.45) and internal temperatures ($T_{int}$ = 1000, 1500, 2000~K). To test hotter and cooler atmospheres, with the latter corresponding to other orbital phases than the day side, we tested models with irradiation temperatures of ($T_{irr}$=1800, 1950, 2050, 2450, 2900, 3100, 3200~K).

In general, the models were difficult to converge because most of the irradiation was in the NUV due to the high effective temperature of the white-dwarf host, resulting in much of the irradiation being absorbed in a relatively small part of the atmosphere. This heating at a relatively discrete layer of the atmosphere results in a sharp temperature inversions at low pressures, a more extreme case of the strong inversions than those seen in ultra-hot Jupiters around early-type main sequence host stars \citep{Lothringer2019}. At some irradiation temperatures, the temperature profile of the atmosphere also crossed important condensation boundaries, when certain short wavelength absorbers will begin to rain out of the atmosphere. This can cause numerical oscillations in the model, where the lower atmosphere will begin to warm without those important upper atmosphere absorbers, causing those same absorbers to evaporate back into the gas phase, and cooling the lower atmosphere through an anti-greenhouse effect. This cycle can continue and cause instabilities in the numerical convergence of these models towards radiative-convective equilibrium.

The morning and evening hemispheres also remaining featureless was somewhat unexpected. Based on the measured brightness temperatures, the terminator regions appeared to be between 1800--2100~K, cool enough to avoid H$_2$O thermal dissociation and for H$^-$ to no longer dominate the opacity. A couple possible scenarios could explain the featureless morning and evening spectra. First, if the brightness temperatures are underestimated by ~500~K, then the morning and evening hemispheres could remain warm enough for H$^-$ opacity to remain dominant.

Alternatively, the morning and evening hemispheres may become increasingly isothermal as the temperature inversion induced by the extreme irradiation on the day side hemisphere begins to wane at longitudes further from the substellar point. Emission spectra in this region would thus remain featureless. One would then expect the undetected night side to finally show evidence of H$_2$O absorption as the temperature structure becomes fully non-inverted. This behavior is qualitatively similar to results from the HST/WFC3/G141 phase curve of ultra-hot Jupiter WASP-121b, which has a nearly identical equilibrium temperature as SDSS\,1557 ($T_{\rm{eq}}$ = 2350~K, \citealt{Mikal-Evans22}).

Another scenario to explain the morning and evening spectra and brightness temperatures would be dilution from the still-visible portion of the day side hemisphere. The portion of SDSS\,1557B visible during the morning and evening phases will include a combination of the illuminated day side and the non-illuminated night side. Thus, the flux measured at the morning and evening phases might be better represented as a combination of a representative day side and night side spectrum, or by diluted day side spectrum \citep{Taylor20}. This latter case is what is effectively done when comparing our model spectra to the normalized observations, rather than comparing absolute fluxes.

\begin{figure}
\begin{center}
\includegraphics[width=0.48\textwidth]{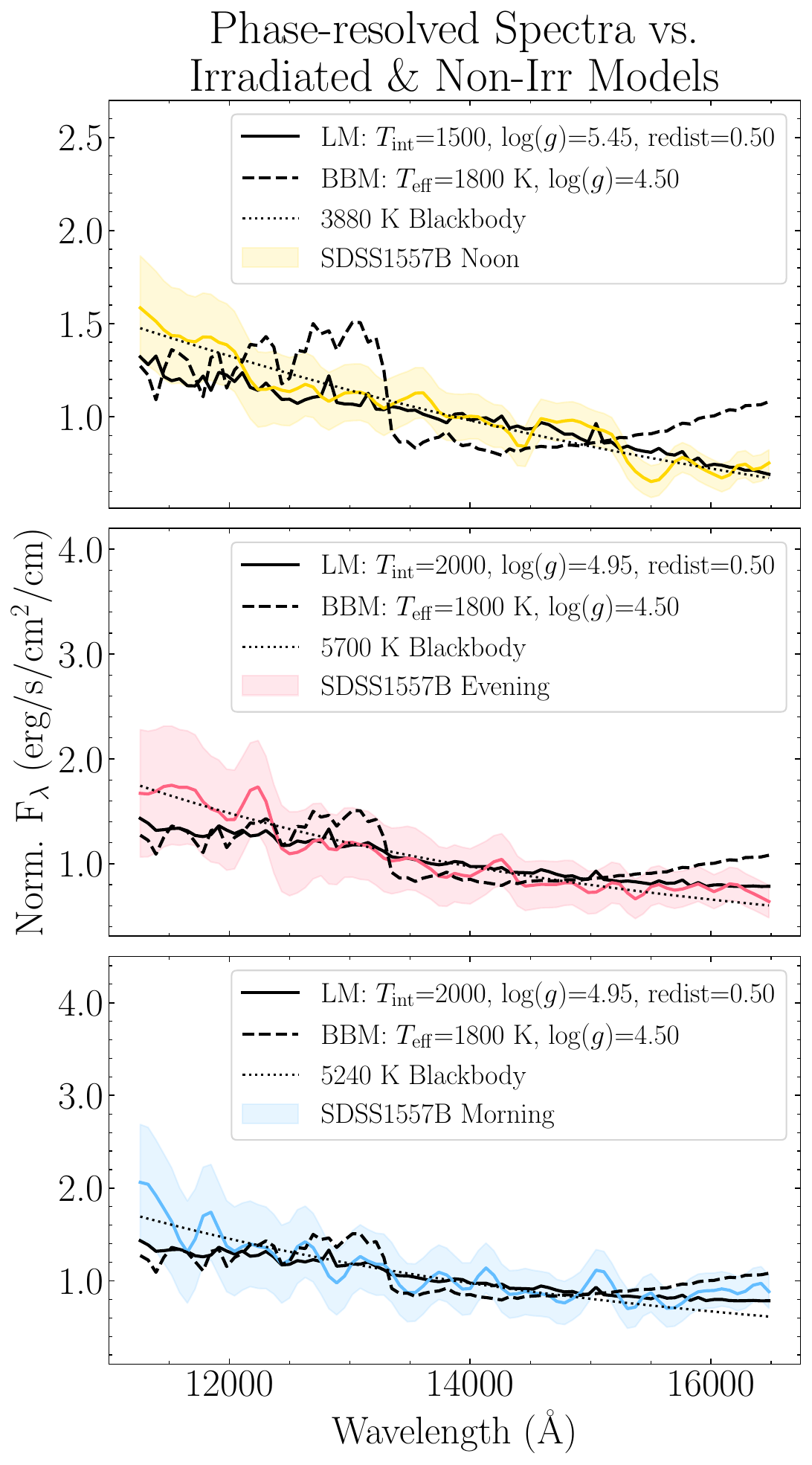}
\caption{Comparison of extracted spectra of SDSS\,1557B with best-fit models from the Lothringer model (LM) grid (Section~\ref{sec:Compare_irr_models}) and the Brock \& Barman Model (BBM) grid (Section~\ref{sec:Compare_nonirr_models}). Best-fitting blackbody curves are also shown. Ignoring Bond Albedo, the irradiation temperature at the substellar point for SDSS\,1557B is 2540~K. Neither model grid is able to match the slope or features of each extracted spectrum. }
\label{fig:modelcomps}
\end{center}
\end{figure}

%%%%%%%%%%%%%%%%%%%%%%%%%%%%%%%%%%%%%%%%%%%
\subsection{Non-Irradiated Models}
\label{sec:Compare_nonirr_models}

It is also illustrative to compare our extracted irradiated BD spectra to models of cloudy, non-irradiated BD atmospheres to get a baseline of what this system may have looked like without its WD primary.  Thus, we conducted a comparative analysis with a set of non-irradiated models specifically designed for the study of L-type BDs \citep{Brock2021}. As with the irradiated models in Section~\ref{sec:Compare_irr_models}, this grid was calculated with PHOENIX, but unlike the irradiated models, assumes local thermodynamic equilibrium. The grid covers $T_{\rm{eff}}$ from 800~K to 2100~K and surface log($g$) from 3.5 to 5.5 (cgs units). Briefly, these models include the effects of non-equilibrium chemistry (parameterized by the eddy diffusion coefficient $K_{\rm{zz}}$) and the effects of cloud opacity. The cloud location, composition and particle number densities are initially determined by equilibrium chemistry.

This initial condensate distribution is then constrained by two free parameters, the pressure at the cloud top and the mean particle size (assuming a log-normal size distribution). Above the cloud top, condensate number density decreases exponentially, while the cloud base is established by equilibrium chemistry. The grid coarsely samples models with cloud top pressures ($P_c$) ranging from 0.1 to 10 bar and mean grain sizes of 0.25, 0.5 and 1 \mum{}.  The \citet{Brock2021} study was focused on a specific set of BD binaries and, consequently, the grid does not uniformly sample all the parameters. However, for the temperatures estimated above, the grid provides good coverage of the expected $T_{\rm{eff}}$ and surface gravity and offers a variety of cloud scenarios to explore.

In Figure~\ref{fig:modelcomps}, the best-fitting cloudy, non-irradiated models for each hemisphere integrated spectrum are labeled as BBM (for Brock and Barman Model). For each phase, Noon, Evening, and Morning, the same non-irradiation model was the closest match. As with the irradiated models, the observed data showcases a slope that is steeper than anticipated by the cloudy, non-irradiated atmosphere models. Our extracted spectra also appears more featureless, as one might expect from the constant irradiation of the primary.

\begin{figure}
\begin{center}
\includegraphics[width=0.48\textwidth]{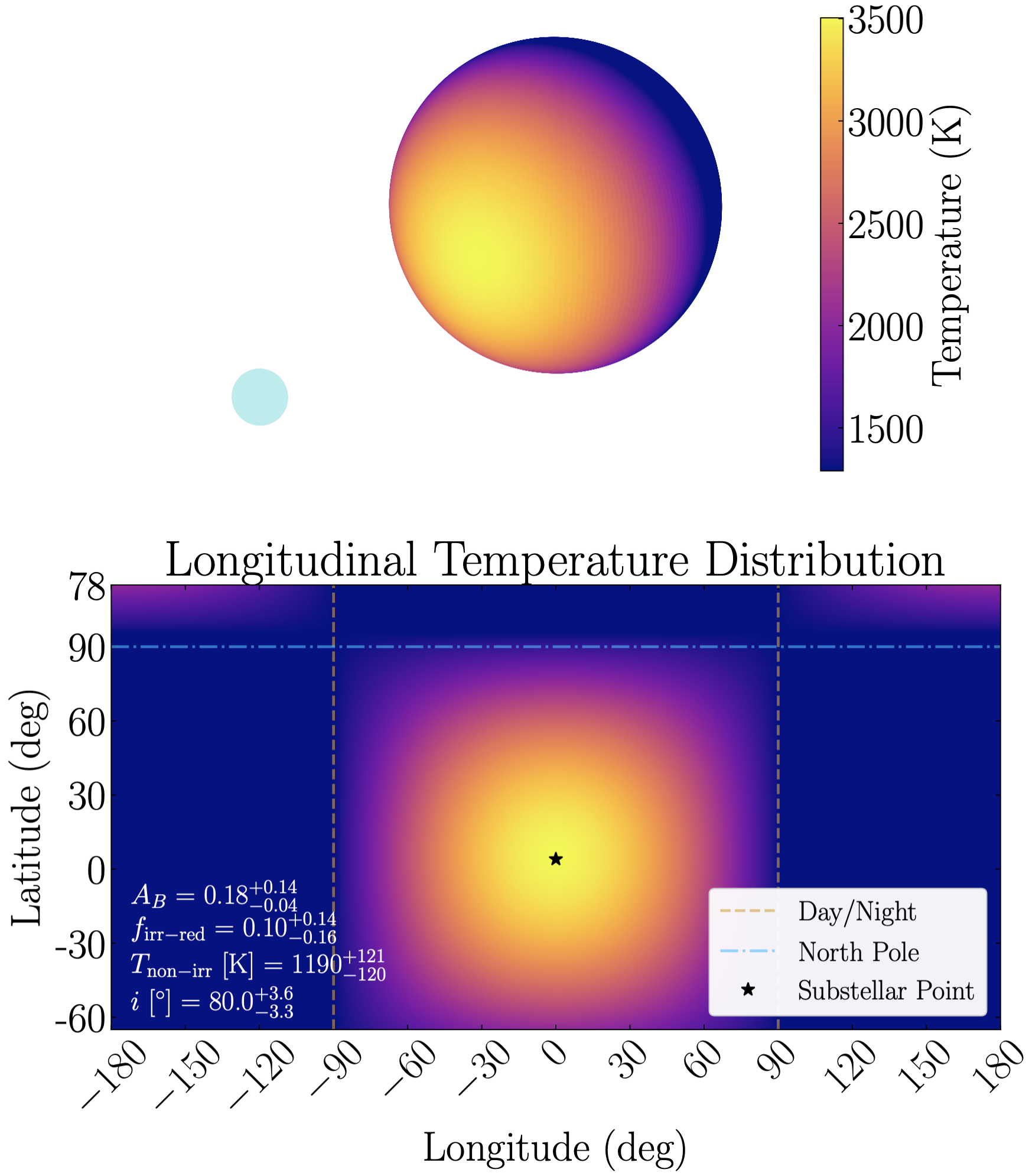}
\caption{\textit{Top}: A 2D depiction of the temperature distribution on SDSS\,1557B projected onto a 3D sphere. Relative radii of WD and BD are to scale, whereas orbital separation is zoomed in for clarity. The WD primary, SDSS\,1557A, is represented as a pale blue dot in the lower left. \textit{Bottom}: A 2D projection of the toy model above with labeled best-fit values and 1$\sigma$ uncertainties for Bond albedo, external irradiation redistribution fraction, non-irradiated BD temperature, and inclination.}
\label{fig:longtempdist}
\end{center}
\end{figure}

%%%%%%%%%%%%%%%%%%%%%%%%%%%%%%%%%%%%%%%%%%%%%%%%%%%%%%%%%%%%%%%%%%%%
%%%%%%%%%%%%%%%%%%%%%%%%%%%%%%%%%%%%%%%%%%%%%%%%%%%%%%%%%%%%%%%%%%%%
\section{Longitudinal Temperature Distribution} \label{sec:tempdist}

One of the key atmospheric processes affecting the structure of tidally-locked irradiated atmospheres is the irradiation redistribution fraction, or efficiency of the heat redistribution from day to night. The strength of this fraction is thought to be determined by a competition between radiative losses and advective heat transport \citep{ShowmanGuillot02, PerezBeckerShowman13, KomacekShowman16, Zhang18}. In theory, a hotter planet should be more efficient at losing heat through radiation, effectively lowering the irradiation redistribution fraction, but this might be balanced by a stronger atmospheric circulation. This trend changes for atmospheres with equilibrium temperatures higher than 2300~K, where hydrogen dissociation/recombination should help with day-to-night heat transport \citep{Bell18, TanKomacek19}. In addition, extremely rapid rotation also tends to inhibit day-to-night heat transport, as appropriate for the close-in BD around WD of our case \citep{TanShowman20_rotationWDBDs}. With our phase-resolved spectroscopy of SDSS\,1557B, we utilized a simple atmosphere to constrain the strength of these atmospheric processes.

Using the calculated brightness temperatures and available physical parameters of the system (Table~\ref{tab:keyprops}), we generated a temperature distribution map of SDSS\,1557, meant to constrain four key atmospheric processes that impact the temperature distribution. The construction of this map utilized a grid search involving the following parameters: Bond albedo ($A_B$), which determined the fraction of incident irradiation reflected away from the day side; irradiation redistribution fraction ($f_{\rm{irr-red}}$), which governed how much non-reflected irradiation is taken away from the day side and evenly redistributed throughout the sphere (i.e., 0.0 = no redistribution; 0.5 = half of non-reflected irradiation flux was uniformly redistributed; 1.0 = total redistribution with uniform temperatures across all latitudes and longitudes); non-irradiated BD temperature ($T_{\rm{non-irr}}$), which established the base temperature of the BD without any irradiation; and lastly, Inclination ($i$), which controlled the visibility of one hemisphere when the other hemisphere dominated the emission. In our set up, we assumed alignment between the BD's rotation inclination and the orbital inclination. This process was also deployed for NLTT5306B \citep{Amaro23}.

Here, we briefly explain the steps involved in constructing this map. First, we created a sphere of uniform temperature, determined by $T_{\rm{non-iir}}$, meant to represent the BD temperature in the absence of external irradiation. We then introduced a source of external irradiation, in this case SDSS\,1557A, where the irradiation strength was governed by $T_{\rm{WD}}$, orbital separation $a$, and $R_{\rm{WD}}$. Immediately, a fraction of this external irradiation was reflected away, a consequence of $A_B$. At this point, day side temperatures were combination of $T_{\rm{non-irr}}$, $A_B$, and external irradiation from the WD primary. Night side temperatures are still set at the initial $T_{\rm{non-irr}}$ value. Quantitavely, day and night temperatures at this step are determined by:
\begin{align*}
    T_{eq}^{\rm{Day}} &= \Big( \frac{L_{\rm{WD}} (1-A_B)}{16 \pi a^2 \sigma_{\rm{SB}}  f_{\rm{rad}} } \Big)^{\frac{1}{4}} \\
    T_{eq}^{\rm{Night}} &= T_{\rm{non-irr}},
\end{align*}
where $L_{\rm{WD}} = 4 \pi R_{\rm{WD}}^2 \sigma_{\rm{SB}} T_{\rm{WD}}^4$, $\sigma_{\rm{SB}}$ is the Stefan-Boltzmann constant, and $f_{\rm{rad}}$ is the fraction of the surface that reradiates (set to 1.0 in our model).

Then, we parameterized the global circulation -- driven by the day-night temperature difference -- with the parameter $f_{\rm{irr-red}}$, which represents the fraction of energy transported from the day side to the night side. In our simple model, we subtracted a percentage of the day side temperatures from each day side cell, summed them, and then redistributed this flux across the whole sphere. Essentially, this acted to decrease the day side temperatures while simultaneously increasing night side temperatures and maintaining a smooth boundary at the day-to-night terminators. Finally, we applied an inclination to the toy atmosphere. To determine the best-fit combination of parameters, the average temperature of each hemisphere ($T_M^H$) was then compared to the respective brightness temperature ($T_B^H$) via the following:
\begin{equation}
    x^H = \Bigg( \frac{T_M^H - T_B^H}{\sigma^H_{T_B}} \Bigg)^2
\end{equation}
where $\sigma^H_{T_B}$ is the uncertainty in the brightness temperature.

Without a detectable midnight brightness temperature, we considered phases ranging from the morning through the noon to the evening to constrain the temperature distribution. Since our current model is symmetric around the substellar point, we fit our model to the average of the morning and evening brightness temperatures in the appropriate phases. However, since morning and evening hemispheres also include half of the day side hemisphere, we decreased the weight of these phases in our fits by 50\%. The final best-fit combination of parameters was thus decided by the lowest value of $x^{noon}$ + ($0.5\times x^{morn}$), where $x^{morn}$ is interchangeable with $x^{eve}$ in our symmetric atmosphere model. 

To quantify uncertainties, we initially marginalized the distributions of each parameter into a one-dimensional array. Then, to establish the 1$\sigma$ range, we added 1 to the lowest value within the array and determined the two points at which this adjustment takes place. These became the upper and lower 1$\sigma$ values.

Figure~\ref{fig:longtempdist} shows the 2D longitudinal temperature distribution with best-fit values labeled in the lower left: $A_B$ = 0.18$^{+0.14}_{-0.04}$, $f_{\rm{irr_red}}$ = 0.10$^{+0.14}_{-0.16}$, $T_{\rm{non-irr}}$ = 1190$^{+121}_{-120}$, $i$ = 80.0$^{+3.6}_{-3.3}$. Given the high day side brightness temperatures and likely regions of molecular dissociation near the substellar point, the low Bond albedo is physically reasonable within its uncertainties.

The low irradiation distribution fraction is also consistent with predictions for strongly irradiated atmospheres, both for hot Jupiters \citep{Perna12, Perez-Becker13, Heng_BriceOlivier13}, and rapidly rotating (assuming tidal locking) irradiated BDs in which fast rotation inhibit day-to-night heat transport \citep{TanShowman20_rotationWDBDs,Lee20, Lothringer_Casewell20, Lee22_sunbathing}.

The best-fit inclination of $i=80^{+3.6}_{-3.3}$ degrees is higher than the previously published value of 62$\pm$3 degrees in \citet{Farihi17}. While the two inclination values are within 3$\sigma$ uncertainty, we believe the higher value found is more accurate, given the higher-quality data and advanced phase-resolved spectroscopy. The enhanced precision afforded by our methodology leads us to have more confidence in the reliability of our result. Additionally, an inclination of 80\% is still physically possible, given the orbital geometry of the system.

%%%%%%%%%%%%%%%%%%%%%%%%%%%%%%%%%%%%%%%%%%%%%%%%%%%%%%%%%%%%%%%%%%%%
\section{Future Observations} \label{sec:future_modeling}
Due to their temperatures, the spectra of cool brown dwarfs and many hot Jupiters peak in the near-infrared (typically between 1--3~$\mu$m). Therefore, our near-infrared observations are particularly constraining for BD atmosphere models. The work described here provided insights into key atmospheric processes within ultracool atmospheres (e.g., vertical mixing, circulation, photochemistry, and condensate clouds). Additionally, through some mismatches between the state-of-the-art models and the observations, we identified some of the current limitations of brown atmospheric models.

With its superior infrared sensitivity and higher spectral resolution, the James Webb Space Telescope (JWST) could offer complementary infrared observations of these rare systems. Already, JWST has enabled orders-of-magnitude increases in the information content of atmospheres for hot Jupiters \citep{Bell23_NatAstro, Mikal-Evans23} and BDs \citep{Miles23, Manjavacas24, Lew24}. Recently, our team was awarded a Cycle 3 JWST proposal that will observe five rare WD--BD systems (JWST GO-4967; PI: Y. Zhou). These observations will reveal processes at a much greater range of pressures, allowing for a more comprehensive comparison between observations and models. In the case of SDSS\,1557, phase-resolved spectroscopy from JWST will enable tighter longitudinal constraints on cloud coverage and possible thermal inversions as well as further characterization of circumbinary debris disk that is actively polluting the primary WD \citep{Farihi17}.

%%%%%%%%%%%%%%%%%%%%%%%%%%%%%%%%%%%%%%%%%%%%%%%%%%%%%%%%%%%%%%%%%%%%
%%%%%%%%%%%%%%%%%%%%%%%%%%%%%%%%%%%%%%%%%%%%%%%%%%%%%%%%%%%%%%%%%%%%
\section{Conclusions}
\label{sec:conclusions}

In this study, we presented time-resolved, spectrophotometric phase-mapping of the highly irradiated BD SDSS\,1557. The key findings of our study are as follows:

\begin{itemize}
    \item We presented high-quality \textit{HST}/WFC3/G141 phase-resolved spectra of WD--BD binary SDSS\,1557. The observations sampled 100\% of the rotational phase of the BD and exhibited strong broadband photometric variations around 21\% from peak-to-trough.
    
    \item We modeled the broadband light curve of SDSS\,1557 to within 2.5\%. Best-fit model consisted of $k$ = 1 and 2 terms with corresponding amplitudes of $amp_1$ = 0.105 $\pm$ 0.001 and $amp_2$ = 0.015 $\pm$ 0.001.
    
    \item Sub--band phase curves made in the $J$--, Water--, and $H'$--bands resulted in peak-to-trough amplitudes of 0.084 $\pm$ 0.001, 0.126 $\pm$ 0.002, and 0.133 $\pm$ 0.002. No relative phase shift between sub--band filters was observed.
    
    \item We reveal a strong wavelength dependence on amplitude using spectroscopically-resolved phase curves, where the amplitudes of both the $k$=1 and $k$=2 terms generally increase with wavelength. The highest amplitude for both terms occurs around 1.50 \mum{}.
    
    \item After modeling and subtracting the WD contribution, we extracted phase-resolved spectra of the irradiated BD SDSS\,1557B. Due to inclination, we estimated up to 15\% of the day side flux is visible from the night side. After subtracting 15\% of the day side from the night side, the midnight phase spectrum was nearly undetectable.
    
    \item The hemisphere averaged brightness temperatures for Noon, Morning, and Evening phases exhibit wavelength dependence, with temperatures that monotonically decrease with wavelength. Upper limits for the midnight phase suggest brightness temperatures no greater than 1600~K.
    
    \item A radiative and energy redistribution model closely reproduced both the observed phase-resolved hemisphere integrated brightness temperatures, excluding the midnight spectrum due to low signal-to-noise. The model suggested an $A_B$ = 0.18$^{+0.14}_{-0.04}$, $f_{\rm{irr-red}}$ = 0.10$^{+0.14}_{-0.16}$, $T_{\rm{non-irr}}$ = 1190$^{+121}_{-120}$ K, and $i$ = 80.0$^{+3.6}_{-3.3}$ for SDSS\,1557B.

    \item The noon-phase spectrum of SDSS\,1557B is well-approximated by the hottest models computed. However, best-fitting models for the morning and evening phases failed to capture the steeper spectral slopes and lack of spectral features. This could be explained by an upper atmosphere that becomes increasingly isothermal at longitudes farther from the equator, as well as the morning and evening phases being diluted versions of the day side hemisphere. 
    
    \item Phase-resolved spectroscopy from JWST will likely capture the night side hemisphere, providing a more complete map of the atmosphere and more precise explanations for the morning and evening spectra of SDSS\,1557B.
    
    \item Overall, this irradiated BD highlights a parameter space that is not yet accurately captured by existing 1D models for highly irradiated, ultracool atmospheres.

\end{itemize}

\section*{acknowledgments}
This material is based upon work supported by the National Science Foundation Graduate Research Fellowship under Grant No. DGE-1746060. Any opinion, findings, and conclusions or recommendations expressed in this material are those of the authors(s) and do not necessarily reflect the views of the National Science Foundation. HST data presented in this paper were obtained from the Mikulski Archive for Space Telescopes (MAST) at the Space Telescope Science Institute. The specific observations analyzed can be accessed via \dataset[10.17909/hbp0-za27]{https://doi.org/10.17909/hbp0-za27}. Support for Program number HST-GO-15947 was provided by NASA through a grant from the Space Telescope Science Institute, which is operated by the Association of Universities for Research in Astronomy, Incorporated, under NASA contract NAS5-26555. SLC acknowledges support from an STFC Ernest Rutherford Fellowship ST/R003726/1. This work has made use of data from the European Space Agency (ESA) mission Gaia (\url{https://www.cosmos.esa.int/gaia}), processed by the Gaia Data Processing and Analysis Consortium (DPAC, \url{https://www.cosmos.esa.int/web/ gaia/dpac/consortium}). Funding for the DPAC has been provided by national institutions, in particular the institutions participating in the Gaia Multilateral Agreement.  We acknowledge the effort of Dr. Siyi Xu, who helped create the approved proposal for these exciting systems and assisted with planning observations. We also thank Professor Jay Farihi for help with clarifying questions during analysis.

%\end{acknowledgments}

\vspace{0.5 cm}
\textit{Author contributions}: D.A. initiated and developed the research project. R.C.A. performed the data reduction, analysis, and drafted the manuscript. D.A. made major contributions to the manuscript. Y.Z. and B.L. provided guidance and and input regarding data reduction and light curve model fitting. J.L. adapted existing 1D atmosphere models to our data and made contributions to the modeling sections of the manuscript. S.C. and X.T. provided guidance regarding the intellectual direction of analysis. T.B. provided existing atmosphere models for comparison with our data. All authors, including L.M., M.M., and V.P., contributed guidance and suggestions to improve the manuscript.

\facility{HST (WFC3)}
%\facilities{}

\software{astropy \citep{astropy2013,astropy2018},  
          Source Extractor \citep{Bertin96_sextractor},
          NumPy \citep{harris2020array},
          SciPy \citep{scipy2020-NMeth},
          Matplotlib \citep{Hunter07_matplotlib},
          HSTaXe (\url{https://github.com/spacetelescope/hstaxe})
          }

%%%%%%%%%%%%%%%%%%%%%%%%%%%%%%%%%%%%%%%%%%%%%%%%%%%%%%%%%%%%%%%%%%%%
%%%%%%%%%%%%%%%%%%%%%%%%%%%%%%%%%%%%%%%%%%%%%%%%%%%%%%%%%%%%%%%%%%%%
\appendix
\section{Updated System Parameters} \label{sec:updateparams}

\textbf{WD Radius}: In order to scale the flux of the WD model to the expected flux at the distance and radius of SDSS\,1557A, we needed an estimate for WD radius. We solved the bolometric luminosity equation for radius:
\begin{equation}
    L_{\rm{Bol}} = 4\pi R_{\rm{WD}}^2 \sigma_{\rm{SB}} T_{\rm{eff}}^4,
\end{equation}

\noindent where $T_{\rm{eff}}$ is the published effective temperature of 21800 $\pm$ 800 K and $\sigma_{\rm{SB}}$ is the Stefan-Boltzmann constant. We determined $L_{\rm{Bol}}$ from $M_{\rm{Bol}}$, the bolometric magnitude. like so:
\begin{equation}
    M_{\rm{Bol}} = BC_X + M_X,
\end{equation}

\noindent where X is the specific photometric filter.

From a literature search, we found a bolometric correction for WDs in the Johnsons-Cousins \textit{V}--band \citep[][]{Carrasco14}. Using data from Table~3 in \citet{Carrasco14}, we created a grid of \textit{V}--band bolometric corrections, and bicubicly interpolated along $T_{\rm{eff}}$ and log($g$). Using this method, we calculate a BC$_{\rm{V}}$ = -2.2384 for SDSS\,1557A. 

With an estimate for bolometric correction, we then needed an absolute magnitude in the \textit{V}--band, which can be calculated from apparent magnitude with:
\begin{equation}
    M_{\rm{V}} = m_{\rm{V}} + 5~\rm{log}_{10}\Big(\frac{D}{10~\rm{pc}}\Big)
\end{equation}

where $D$ is the distance. For this variable, we adopted the $\textit{Gaia}$ photogeometric distance of $D$ = 500$^{+19.8}_{-18.0}$ pc from \citet{BailerJones21}. For \textit{V}--band apparent magnitude, we use a photometric transformation\footnote{Equation for photometric transformation comes from Tables 5.7 and 5.9 in Section 5.5.1 of the Gaia DR3 documentation, by Josep Manel Carrasco and Michele Bellazzini.} from $\textit{Gaia}$ $G$, $GBP$, and $GRP$ filters to Johnson-Cousins filters. Specifically, we used the equation:
\begin{equation}
    G-V = c_0 + c_1(GBP-GRP) - c_2(GBP-GRP)^2 + c_3(GBP-GRP)^3
\end{equation}
where c$_0$ = -0.02704, c$_1$ = 0.01424, c$_2$ = 0.2156, and c$_3$ = 0.01426. With apparent $\textit{Gaia}$ magnitudes of $G$ = 18.609 $\pm$ 0.004, $GBP$ = 18.519 $\pm$ 0.024, $GRP$ = 18.810 $\pm$ 0.042, we solved for V, yielding an apparent \textit{V}--band magnitude of m$_V$ = 18.659 $\pm$ 0.008 for SDSS\,1557A.

Thus, we calculated an absolute \textit{V}--band magnitude of M$_V$ = 10.164 $\pm$ 0.241. Combining M$_V$ with BC$_V$, we calculate a bolometric magnitude of $M_{\rm{Bol}}$ = 7.926 $\pm$ 0.241, which translates to a bolometric luminosity of $L_{\rm{Bol}}$ = 2.035$\times$10$^{32}$ $\pm$ 3.057$\times$10$^{28}$. Finally, we solved for the radius of the WD: $R_{\rm{WD}}$ = 0.0162 $\pm$ 0.0012 R$_{\odot}$.

\textbf{BD Radius}: We estimate the radius of the irradiated BD SDSS\,1557B using evolutionary models from the Sonora Bobcat model grid \citep{Marley21_Sonora21}. The evolutionary portion predicts how parameters such as mass and radius evolve over time. For our system, we used the evolutionary tables with solar metallicity ([M/H]=0.0), given the compelling evidence for ongoing accretion from a debris disk that lies outside of the system \citep{Farihi17}. 

Similar to the method used in \citet{Beiler23}, we randomly draw a million pairs of mass and age values, then interpolate across the models for each pair to generate a range of possible radii. For the input masses and ages, we chose uniform distributions with the following boundary conditions: 60 $< M_{\rm{BD}} <$ 75 $\rm{M_{Jup}}$ and 0.03 $<$ Age$_{\rm{BD}}$ $<$ 10 Gyr. We adopted the resulting range of BD radii, $R_{\rm{BD}}$ = 0.106$\pm$0.024 R$_{\odot}$, as the radius of SDSS\,1557B (see Table~\ref{tab:keyprops}).

\textbf{Distance}: Original value from \citet{Farihi17} was 520$\pm$35 pc. \citet{Farihi17} estimated WD parameters $T_{\rm{eff}}$ and log($g$) from fitting atmospheric models to Balmer lines in 60 individual XSHOOTER spectra. Published errors for those parameters were the standard deviation from all 60 measurements. Then, errors in derived stellar parameters, like distance, were calculated by propagating the uncertainties in the adopted $T_{\rm{eff}}$, log($g$), and published photometry through WD evolutionary models \citep[][]{Fontaine01}.

We adopted an updated distance value, $D$ = 500$^{+19.8}_{-18.0}$ pc, from the \citet{BailerJones21} distance catalog for WDs. For most WDs in this catalog, the authors inferred two types of distances: (1) geometric, which uses parallax with a direction-dependent prior on distance, and (2) photogeometric, which additionally uses the color and apparent magnitude of the WD. SDSS\,1557 had both distances published, with $D$ = 590.1$^{+74.3}_{-64.6}$ pc as the geometric distance, and $D$ = 500.0$^{+19.8}_{-18.0}$ pc as the photogeometric distance. We adopted the photogeometric distance in our study, as this is a more reliable estimate.
\restartappendixnumbering

\begin{figure*}
\begin{center}
\includegraphics[width=0.78\textwidth]{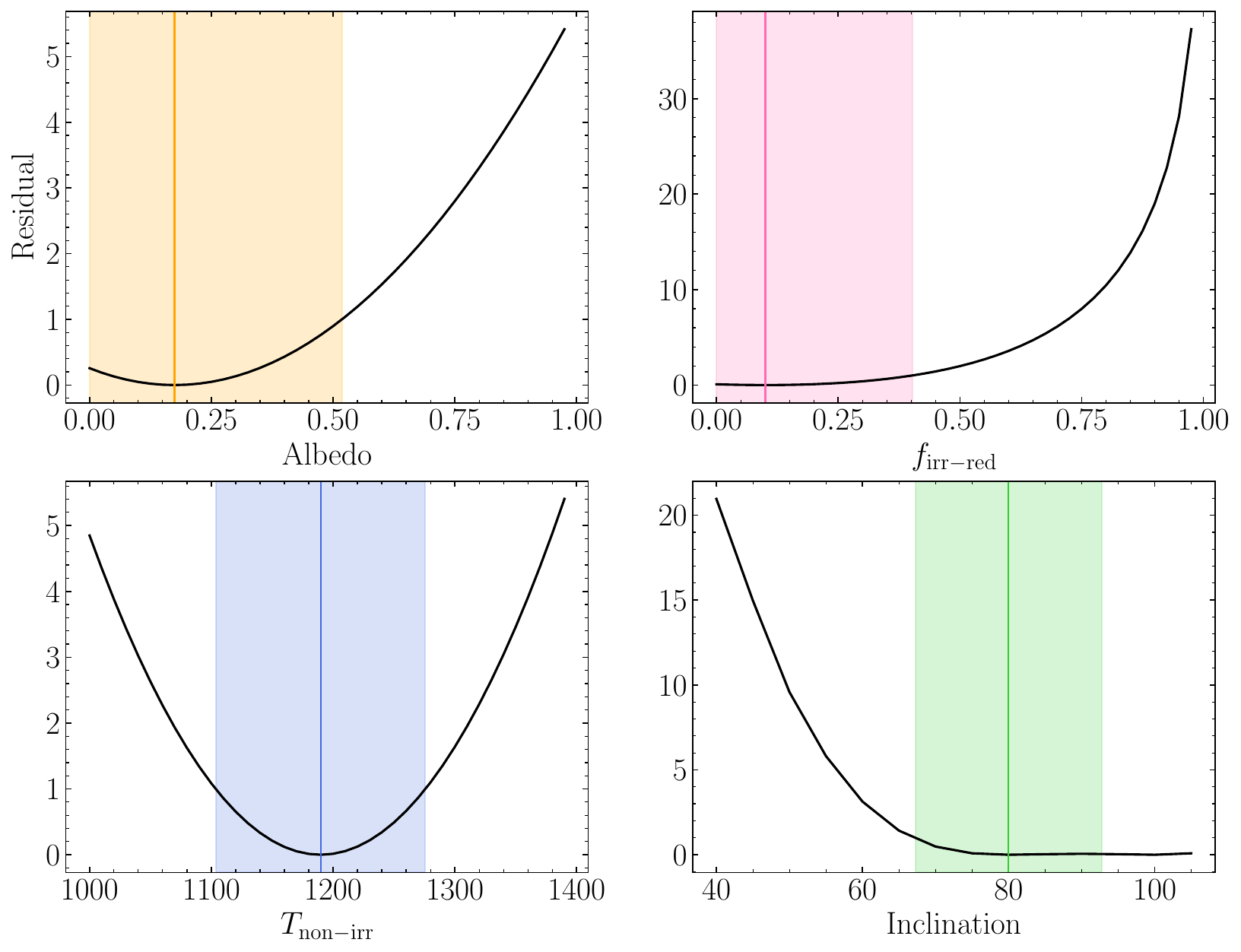}
\caption{Residual curves for each parameter explored in the simple heat redistribution atmosphere model shown in Figure~\ref{fig:longtempdist}. Best-fit values, which are the lowest points in the curves, are highlighted by the vertical line in each panel, while 1$\sigma$ uncertainties are shown as the shaded regions.  }
\label{fig:bestfit_longtemp}
\end{center}
\end{figure*}

\section{Best-fit Longitudinal Temperature Distribution} \label{sec:longtemp}

We conducted a grid search among 4 parameters: (1) Bond Albedo, $A_{\rm{B}}$=[0.0, 1.0] in increments of 0.025, (2) Irradiation redistribution fraction, $f_{\rm{irr-red}}$=[0.0, 1.0] in increments of 0.025, (3) non-irradiated brown dwarf temperature, $T_{\rm{non-irr}}$=[1000,1400] in increments of 10~K, and (4) Inclination, $i$=[40,110] in increments of 5 degrees. The method for finding the best-fit combination of parameters is largely similar to what is described in Section~C of \citet{Amaro23}, with one key exception. In NLTT5306B, we compared the hemisphere integrated day  and night side temperatures of the simple atmosphere model to the day and night brightness temperatures derived from the extracted BD spectra. For SDSS\,1557B, the extracted midnight spectrum was just below significant detection levels. Thus, we altered the fitting criteria to consider additional phases, e.g. morning and evening, as a replacement for the midnight spectrum. The resulting best-fit curves for each parameter are shown in Figure~\ref{fig:bestfit_longtemp}.

% A handy "cheat sheet" that provides the necessary \latex\ to produce 17 
% different types of tables is available at \url{http://journals.aas.org/authors/aastex/aasguide.html#table_cheat_sheet}.

%%%%%%%%%%%%%%%%%%%%%%%%%%%%%%%%%%%%%%%%%%%%%%%%%%%%%%%%%%%%%%%%%%%%
%%%%%%%%%%%%%%%%%%%%%%%%%%%%%%%%%%%%%%%%%%%%%%%%%%%%%%%%%%%%%%%%%%%%
\bibliography{main}
\bibliographystyle{aasjournal}

\end{document}